\documentclass[12pt]{article}

\usepackage[nosort]{cite}
\usepackage{epsfig}
\usepackage{amsmath}
\usepackage{amsfonts}
\usepackage{amssymb}
\usepackage{multirow}
\usepackage{hyperref}

\newcommand{\dslash}{\not\!\partial}

\begin{document}

\setcounter{page}{0}
\thispagestyle{empty}


\vskip 8pt

\begin{center}
{\bf \Large {
$\mathbf{\Delta F=1}$ constraints on composite Higgs models with LR parity
}}
\end{center}

\vskip 10pt

\begin{center}
{\large Natascia Vignaroli}
\end{center}

\vskip 20pt

\begin{center}

\centerline{ {\it Department of Physics and Astronomy, Iowa State University, Ames, IA 50011, USA}} \centerline{ {\it Department of Physics and Astronomy, Michigan State University, East Lansing, MI 48824, USA}}
\end{center}

\vskip 50pt

\begin{abstract}
\vskip 3pt
\noindent
We analyze the bounds on the spectrum of composite Higgs models (CHM) that come from flavor observables, by means of simple two-site effective Lagrangians, which incorporate a custodial symmetry and a Left-Right parity and which could also be adopted in further phenomenological studies on CHM.\\
We derive, in particular, an important constraint on the masses of the $(t_L, b_L)$ partners, which does not depend on the flavor structure of the sector beyond the SM. This bound is obtained from the ``infrared" contribution to $b \to s\gamma$ induced by the flavor-conserving effective vertex $Wt_Rb_R$.
We find that the presence of a custodial symmetry can play a role in protecting this effective coupling and, as a consequence, in attenuating the constraint, which, however, remains of the order of $1$ TeV.\\ In addition to this bound, we calculate the constraints from the ``ultraviolet" contribution to $b \to s \gamma$, induced by loops of heavy fermions, and to $\epsilon^{'}/\epsilon_K$.
\end{abstract}

\vskip 13pt
\newpage



\section{Introduction}
\label{sec:introduction}
A possible solution to the hierarchy problem is based on an 
analogy with the pion mass stabilization in QCD: 
the Higgs, similarly to the pion, might be a composite state, generated by a new strong dynamics; as such, its mass is not sensitive
to radiative corrections above the compositeness scale, assumed to be of the order of the TeV scale. A further protection, that allows the Higgs
to be naturally lighter than the other resonances, exists if the composite Higgs is also  
the pseudo-Goldstone boson of a spontaneously broken global symmetry \cite{Georgi_Kaplan}. A pseudo-Goldstone boson Higgs is expected to be light and 
as such in agreement with the indication from the LEP electroweak precision data (EWPD).  
In this project we will reconsider the bounds on the spectrum of Composite Higgs Models (CHM) that come from flavor observables, with a special focus to $b\to s \gamma$.
Instead of considering a full theory we will work in an effective description valid at low energy.
In particular, we will refer to a ``two-site'' (TS) description \cite{Sundrum,Servant}, where two sectors, the weakly-coupled sector of the elementary fields
and the composite sector, that comprises the Higgs, are linearly coupled each other through mass mixing terms \cite{Kaplan}. After diagonalization the 
elementary/composite basis rotates to the mass eigenstate one, made up of SM and heavy states that are admixture of elementary and composite modes. 
Heavier particles have larger degrees of compositeness: heavy SM particles, like the top, are more composite while the light ones are almost elementary.
In order for composite Higgs models to be compatible with LEP precision data, the presence of a custodial symmetry in the 
composite sector is strongly suggested to avoid large corrections to the $\rho$ parameter. The absence of large Flavor-Changing Neutral Currents is achieved, instead, by a sort of GIM mechanism, that naturally emerges when the connection between 
the elementary and the strong sector proceeds via linear couplings \cite{RSgim}. In absence of a symmetry protection, the LEP data also point toward a small degree of compositeness of the left-handed
 bottom quark (small corrections to $Z\bar{b}_{L}b_{L}$), and, by gauge invariance, of the left-handed top as well. This implies that, in order to 
obtain a heavy enough top quark, it is necessary to have an almost fully composite right-handed top quark.
It has been shown, however, that the corrections to $Z\bar{b}_{L}b_{L}$ can be suppressed if the custodial symmetry 
of the strong sector includes a Left-Right parity \cite{zbb}. This can allow for a smaller right-handed top compositeness.\\
In order to study the phenomenology at energies lower than the compositeness scale, we derive two different models which incorporate a custodial symmetry and a Left-Right parity. We label such models as TS5 and TS10.  
They describe the low-energy regime of the Minimal Composite Higgs Models (MCHM) defined in Ref. \cite{mchm, mchm2}, in the limit
in which only the leading terms in an expansion in powers of the Higgs field are retained
\footnote{see Ref. \cite{Panico-Wulzer}, for two- and three-site effective theories where the full Higgs non-linearities are
included.}. 
In MCHM the Higgs arises as the pseudo-Goldstone boson associated to the $SO(5)\to O(4)$ breaking in the composite sector; where
$O(4)$ includes $SO(4)\sim SU(2)_L\times SU(2)_R$ as well as a parity $P_{LR}$ which exchanges $SU(2)_L$ with $SU(2)_R$. 
Composite fermions can be embedded in a $5=(2,2)+(1,1)$ representation of $SO(5)$ in the TS5 model and in a $10=(2,2)+(1,3)+(3,1)$ in the TS10. 
TS5 and TS10 extend the two-site description of \cite{Sundrum,Servant} to consider 5 and 10 $SO(5)$ representations for composite fermions. In particular, the TS5 model extends the `two site' model of Ref. \cite{Servant} to include the composite fermions needed to give mass to the bottom quark. \\

We find two important bounds on the masses of the heavy fermions which do not depend on the flavor structure of the sector beyond the SM (BSM).
The first comes from the measurement of the $Zb_L\bar{b}_L$ coupling, that we already mentioned and that can be suppressed assuming a $P_{LR}$ symmetry.
The second is obtained from the infrared (IR) contribution to $b \to s\gamma$ induced by the flavor conserving effective vertex $Wt_Rb_R$. 
In composite Higgs models there are two classes of effects that lead to a shift of the 
$b\rightarrow s\gamma$ decaying rate compared to the SM prediction: 
loops of heavy fermion resonances from the strong sector give a ultraviolet (UV) local contribution; 
they generate, at the compositeness scale, the flavor-violating dipole operators $\mathcal{O}_{7}$ and $\mathcal{O}^{'}_{7}$,
which define the effective Hamiltonian for the $b\rightarrow s\gamma$ decay.
The virtual exchange of heavy resonances also generates
the effective V+A interaction of the $W$ boson and the SM quarks, $Wt_R b_R$, which in turn leads to
a shift to $b\to s\gamma$ via a loop of SM particles. This latter IR contribution is enhanced by a chiral factor $m_t/m_b$ and, since in
this case the flavor violation comes entirely from the SM V-A current, $\bar{t}_L\gamma^{\mu}s_L$, it gives a Minimal Flavor Violating (MFV) lower bound on the heavy fermion masses. \\
We also discuss the role of a parity $P_C$, which is a subgroup of the custodial $SU(2)_V$, to protect the effective coupling $Wb_R t_R$.\\ 
In general, stronger bounds can be obtained from the UV CHM contribution to $b\to s \gamma$ and from $\epsilon^{'}/\epsilon_K$ \cite{Isidori}; however, these latter bounds are model dependent and in principle could be loosened by acting on the NP flavor structure (see, for example, \cite{Redi_Weiler}).
The bound from the IR contribution to $b\to s \gamma$, on the other hand, is robust, since it is a MFV effect. \\

The paper is organized as follows: in sec. \ref{sec:model} we introduce our two-site models; in sec. \ref{sec:bound} we discuss the bound from $b \to s\gamma$; we first calculate the MFV bounds from the infrared contribution in generic CHM, by NDA, and in the specific TS5 and TS10, we then proceed to calculate the non-MFV constraints from $b\to s \gamma$ and from $\epsilon^{'}/\epsilon_K$; we draw our conclusions in sec. \ref{sec:conclusions}.

\section{Effective theories for composite Higgs models}
\label{sec:model}
The idea behind Composite Higgs Models is that the Electro Weak Symmetry Breaking may be triggered by a new strong dynamics, 
in analogy with the chiral symmetry breaking in QCD.
In these theories a new strong sector couples to a weakly coupled sector, which coincides with that of the Standard Model without the Higgs. 
The Higgs, as the pion in QCD, is a composite state coming from the latter strong dynamics. Its composite nature  
allows for a solution to the hierarchy problem. Indeed, its mass is not sensitive
to radiative corrections above the compositeness scale, assumed to be of the order of the TeV. 
The EWSB is transmitted to SM fermions by means of linear couplings \cite{Kaplan} (generated by some UV physics at the UV scale $\Lambda_{UV}$) 
between elementary fermions $\psi$ and composite fermions
\begin{equation}
\Delta\mathcal{L}=\lambda\bar{\psi}\mathcal{O}+h.c. 
\label{linCOUP}
\end{equation} 
This way to communicate the EWSB can give a natural explanation of the hierarchies in the quark
masses (through RG evolution of the composite-elementary couplings $\lambda_i$) 
and avoid the tension which occurs when trying to generate large enough quark masses and, at the same time, suppressing FCNC processes\footnote{Tension that instead affects Technicolor and Extended Technicolor Models.}. \\                                                                                                         
As a consequence of linear couplings a scenario of \textit{Partial Compositeness} of the SM particles emerges.
At energies below the compositeness scale a composite operator $\mathcal{O}$ can excite from the vacuum a tower of composite fermions of increasing mass.
Linear couplings (\ref{linCOUP}) thus turn into mass mixing terms between elementary fermions and towers of composite fermions $\chi_n$
\begin{equation}
 \langle 0|\mathcal{O}|\chi_n\rangle=\Delta_n\ \,\ \ \ \mathcal{L}_{mix}=\sum_n\Delta_n \left(\bar{\psi}\chi_n+h.c.\right )\ .
\label{mixTERM}
\end{equation} 
\begin{equation}
 \mathcal{L}=\mathcal{L}_{el}+\mathcal{L}_{com}+\mathcal{L}_{mix} 
\label{Lfull}
\end{equation} 
Because of the mass mixing terms the physical eigenstates, made up of SM and (new) heavy states, are admixture of elementary and composite modes.\\ 
The low-energy phenomenology of such theories can be exhaustively studied, and calculation can be made easier, by considering a truncation of each tower of composite fermions to the first resonance, while 
other heavy states are neglected \cite{Sundrum}. For example, the effective Lagrangian describing one elementary
chiral field $\psi_L$ and its composite partner $\chi$ is
\begin{equation}
 \Delta\mathcal{L}=\bar{\psi}_L i{\not}\partial \psi_L+\bar{\chi} (i{\not}\partial-m_*)\chi- \Delta_L \bar{\psi}_L\chi_R + h.c. \ .
\label{Lchiral}
\end{equation} 
We can rotate the fermions from the elementary/composite basis to the mass eigenstate one, the light(SM)/heavy basis, according to:
\begin{align}
\begin{split}\label{ele/compROTATION}
& \tan\varphi_{L}=\frac{\Delta_{L}}{m_*} \ \ \ \ 
	 \left\{ \begin{array}{l} 
	| light\rangle= \cos\varphi_L |\psi_L\rangle -\sin\varphi_L |\chi_L\rangle \\
	| heavy\rangle= \sin\varphi_L |\psi_L\rangle +\cos\varphi_L |\chi_L \rangle \end{array} \right.
\end{split}
\end{align}
Our eigenstate fields are thus a heavy fermion of mass $m=\sqrt{m^2_* + \Delta^2_L}$ and a light fermion, to be identified with the SM field, that will
acquire a mass after the EWSB. These fields, as we see, are superpositions of elementary and composite states. The angle $\varphi_L$ parametrizes the degree
of compositeness of the physical fields. In particular, the SM fermion has a $\sin\varphi_L\equiv\frac{\Delta_L}{\sqrt{m^2_* + \Delta^2_L}}$ degree of compositeness 
(and a $\cos\varphi_L\equiv\frac{m_*}{\sqrt{m^2_* + \Delta^2_L}}$ degree of elementarity); the mass mixing parameter $\Delta_L$ can be naturally much smaller than the mass 
$m_*$ of the composite fermion\footnote{As a result of RG evolution above the compositeness scale. 
The smallness of $\Delta$ parameters also allows for a sort of GIM mechanism that suppresses large Flavor-Changing Neutral Currents \cite{RSgim}.},
 therefore, SM fermions are in general mostly elementary with a small degree of compositeness, while heavy fermions are mostly composite with a small degree
of elementarity. We have a similar rotation, with angle $\varphi_R$, in the case of right-handed fermions.
SM fermions acquire a mass after the EWSB; since the origin of this breaking resides, by assumption, in the composite sector 
(the Higgs is a fully composite state), the SM fermion mass arises from the composite part of left-handed and right-handed SM fields:
\begin{equation}
 m_\psi = Y_* \frac{v}{\sqrt{2}} \sin\varphi_L \sin\varphi_R ,
\label{mass}
\end{equation} 
where $Y_*$ is a Yukawa coupling among composites, from which the SM Yukawa $y= Y_* \sin\varphi_L \sin\varphi_R$ originates. 
In the following we will assume that the strong sector is flavor
anarchic, so that there is no
large hierarchy between elements within each matrix $Y_*$ and the hierarchy in the masses and mixings of the SM quarks 
comes entirely from the hierarchy in the elementary/composite mixing angles 
(such `anarchic scenario' has been extensively
studied in the framework of 5D warped models, see Refs. \cite{RSgim,Huber,Csaki_anarchic,Casagrande,Albrecht}).
From (\ref{mass}) we can see that heavier SM particles have larger degrees of compositeness: 
heavy SM particles, like the top, have to be quite composite while the light ones are almost elementary.\\

Experimental data give hints on the type of the new strong dynamics responsible for the EWSB. 
The LEP precision data suggest the presence of a custodial symmetry in the 
composite sector to avoid large corrections to the $\rho$ parameter. 
In order to protect $\rho$ (or equivalently the Peskin-Takeuchi T parameter) the composite sector must respect, minimally, a global symmetry:
\[
 SU(2)_L \times SU(2)_R \times U(1)_X\ ,
\]
where $SU(2)_L\times SU(2)_R$ is broken to the diagonal $SU(2)_V$ after the EWSB; 
the unbroken $SU(2)_V$ invariance acts as a custodial symmetry so that $\rho=1$ at tree level.\\
The SM electroweak group $SU(2)_{L}\times U(1)_{Y}$ can be embedded into $SU(2)_{L}\times SU(2)_{R}\times U(1)_{X}$, so that
hypercharge is realized as $Y = T^3_{R} +X$. 
The Composite Higgs transforms as a bidoublet $(2,2)$ under $SU(2)_{L}\times SU(2)_{R}$, $\mathcal{H}\equiv (H,H^c)$, 
where $H$ is the Composite Higgs doublet and $H^c=i\sigma^2 H^*$ is its conjugate.
The $\mathcal{H}$ VEV 
breaks the $SU(2)_{L}\times SU(2)_{R}\times U(1)_{X}$ group down to $SU(2)_{V}\times U(1)_{X}$ and leads to the EWSB. 
Therefore, we have the following relation among charges:
\begin{equation}
	Q=T^3_{L}+T^3_{R}+X=T^3_{L}+Y \ .
	\label{eq.charge}
\end{equation}

This scheme can also results from models where the Higgs arises as the pseudo-Goldstone boson associated to a $SO(5)\to SO(4)\sim SU(2)_L\times SU(2)_R$ breaking in the composite sector;
or to a $SO(5)\to O(4)$ breaking,
 where $O(4)$ includes $SO(4)\sim SU(2)_L\times SU(2)_R$ as well as a parity $P_{LR}$ which exchanges $SU(2)_L$ with $SU(2)_R$. 
This enhanced custodial symmetry can suppress the corrections to the coupling $Z\bar{b}_{L}b_{L}$, which are strongly constrained by LEP data \cite{zbb}. 

\subsection{$P_{LR}$ and $P_C$ symmetries}\label{PLR}
In MCHM \cite{mchm} the Higgs arises as the pseudo-Goldstone boson associated to the $SO(5)\to O(4)$ breaking in the composite sector; 
where the enhanced custodial symmetry $O(4)$ includes $SO(4)\sim SU(2)_L\times SU(2)_R$ as well as a parity 
$P_{LR}$ which exchanges $SU(2)_L$ with $SU(2)_R$. 
As shown in \cite{zbb}, this $P_{LR}$ parity, as well as the $P_C$ symmetry, subgroup of the custodial $O(4)$,
 can protect the coupling $Z\bar{b}_{L}b_{L}$ against large corrections from the composite sector. \\
Each composite operator has a definite left and right isospin quantum number, $T_{L,R}$,
and a 3rd component, $T^3_{L,R}$. 
We can also univocally assign to each SM field definite quantum
numbers, $T_{L,R}$, $T^3_{L,R}$, corresponding to those of the composite operator to which it couples. 
$P_{LR}$ and $P_C$ are symmetries of the composite sector, $P_{LR}$ exchanges $SU(2)_L$ with $SU(2)_R$ 
and $P_C$ is the subgroup of $SU(2)_V$ that transforms $\left |T_L, T_R; T^3_L, T^3_R\right\rangle \to \left |T_L, T_R; -T^3_L, -T^3_R\right\rangle$ 
($SO(3)$ vectors transform with $P_C=diag(1,-1,-1)$).
For $P_{LR}$ ($P_C$) to be a symmetry also of the interacting terms between SM fields and composite operators, 
$\Delta\mathcal{L}=\lambda\bar{\psi}\mathcal{O}+h.c.$, the SM fields $\psi$ have to be eigenstates of $P_{LR}$ ($P_C$). 
This implies:
\begin{equation}
 T_L=T_R \ \ (T^3_L=T^3_R) \ \ (P_{LR}\ invariance)
\end{equation}
\begin{equation}
 T^3_L=T^3_R=0 \ \ (P_{C}\ invariance)\ .
\end{equation} 
\noindent
If the above formulas hold, we can see that the coupling $Z\psi\bar{\psi}$, 
\begin{equation}
g_{\psi}=\frac{g}{\cos\theta_W}(Q^3_L-Q\sin^2\theta_W)\ ,
\end{equation}
 is protected against large corrections. 
Indeed, the electric charge $Q$ is conserved and the charge of the $SU(2)_L$ 3rd component, $Q^3_L$, 
is conserved by custodial invariance plus $P_{LR}$ symmetry and by $P_C$ symmetry. 
By custodial $U(1)_V$ invariance, $\delta Q^3_V=\delta Q^3_R+\delta Q^3_L=0$; if there is also a $P_{LR}$ invariance, 
$\delta Q^3_R=\delta Q^3_L$, therefore $\delta Q^3_L=0$. The same conservation, $\delta Q^3_L=0$, is obtained by $P_C$ invariance: 
the SM $W^3_L$ has an odd parity under $P_C$, $W^3_L\to -W^3_L$; if $\psi$ is a $P_C$ eigenstate it must have $T^3_L=T^3_R=0$, then 
the current $\bar{\psi}\gamma^{\mu}\psi$ is even under $P_C$ and it cannot couple to $W^3_L$, which is odd.\\
We will show (sec. \ref{PCprotect}) that the $P_C$ symmetry can also protect in a similar way the effective coupling $Wt_R b_R$ and, as a consequence, 
it can be responsible for an attenuation of the bound on heavy fermion masses, coming from the process $b\to s\gamma$. \\

In what follows we present the two-site models, TS5 and TS10, which incorporate a custodial symmetry and a $P_{LR}$ parity.\footnote{The TS5 model has been already briefly described in \cite{Bini}, where it was adopted to study the phenomenology of heavy-colored vectors at the LHC.}

\subsection{TS5}\label{TS5}
In the TS5 model, we consider composite fermions filling the following $SO(4)\times U(1)_{X} \sim SU(2)_{L}\times SU(2)_{R}\times U(1)_{X}$ representations:

\begin{align}\label{eq.fields}
\begin{split}	
&	\mathcal{Q}=\left[\begin{array}{cc}
	T & T_{5/3} \\ 
	B & T_{2/3} \end{array}\right]=\left(2,2\right)_{2/3} \ \tilde{T}=\left(1,1\right)_{2/3}\\
&		\mathcal{Q'}_{-1/3}=\left[\begin{array}{cc}
	B_{-1/3} & T' \\ 
	B_{-4/3} & B' \end{array}\right]=\left(2,2\right)_{-1/3} , \ \tilde{B}=\left(1,1\right)_{-1/3}\\
 \end{split}
\end{align}
\noindent
and the composite Higgs in:
\begin{align}\label{eq.higgs}
\begin{split}	
&		\mathcal{H}=\left[\begin{array}{cc}
	\phi^{\dag}_{0} & \phi^{+} \\
	-\phi^{-} & \phi_{0} \end{array}\right]=\left(2,2\right)_{0} \\	
\end{split}
\end{align}

\noindent
The $SO(4)$ multiplets of composite fermions can be embedded into fundamentals $\mathbf{5_{2/3\ (-1/3)}}$ of $SO(5)\times U(1)_{X}$, that decompose as $\mathbf{5_{2/3\ (-1/3)}}=\mathbf{(2,2)_{2/3\ (-1/3)}}\oplus\mathbf{(1,1)_{2/3\ (-1/3)}}$ under 
$SU(2)_{L}\times SU(2)_{R}\times U(1)_{X}$ (see Ref. \cite{DeCurtis:2011yx} for a study of the same representations in a two-site description of $SO(5)$).
We are thus introducing two classes of composite fermions, 
those filling a $\mathbf{5_{2/3}}$ representation, with $X$ charge $X=2/3$ and those in a $\mathbf{5_{-1/3}}$, with $X=-1/3$. 
We want to consider, indeed, the possibility that the SM quark doublet $(t_{L}, b_{L})$ couples to two different BSM operators,
 $\mathcal{Q}_{2/3}$ and $\mathcal{Q'}_{-1/3}$, the first responsible for generating the top mass, the second for generating the bottom mass. 
$(t_{L}, b_{L})$ is linearly coupled to $(T, B)$ through a mass mixing term we call $\Delta_{L1}$ and to $(T', B')$ through a mass mixing term
$\Delta_{L2}$. $t_{R}$ and $b_{R}$ couple respectively to $\tilde{T}$, through a mass mixing term $\Delta_{R1}$, and to $\tilde{B}$, 
through a mass mixing term $\Delta_{R2}$. The fermionic Lagrangian reads, in the elementary/composite basis:

\begin{align}
\begin{split}\label{eq.Lagrange1}
\mathcal{L}=\  & \bar{q}^i_{L}i\dslash q^i_{L}+\bar{u}^i_{R}i\dslash u^i_{R}+\bar{d}^i_{R}i\dslash d^i_{R}\\ 
& +Tr\left\{\bar{\mathcal{Q}}\left(i\dslash-M_{Q*}\right)\mathcal{Q}\right\}+\bar{\tilde{T}}\left(i\dslash-M_{\tilde{T}*}\right)\tilde{T}+Y_{*U}Tr\left\{\bar{\mathcal{Q}}\mathcal{H}\right\}\tilde{T}\\ 
& +Tr\left\{\bar{\mathcal{Q'}}\left(i\dslash-M_{Q'*}\right)\mathcal{Q'}\right\}+\bar{\tilde{B}}\left(i\dslash-M_{\tilde{B}*}\right)\tilde{B}+Y_{*D}Tr\left\{\bar{\mathcal{Q'}}\mathcal{H}\right\}\tilde{B}\\ 
& -\Delta_{L1}\bar{q}^3_{L}\left(T,B\right)-\Delta_{R1}\bar{t}_{R}\tilde{T}-\Delta_{L2}\bar{q}^3_{L}\left(T',B'\right)-\Delta_{R2}\bar{b}_{R}\tilde{B}+h.c.	\ .
\end{split}
\end{align}
\noindent
where the superscript $i$ runs over the three SM families ($i$ = 1, 2, 3), with $q^3_L\equiv(t_L , b_L )$, 
$u^3\equiv t_R$, $d^3\equiv b_R$. By construction, the elementary fields couple to the composite
ones only through the mass mixing terms, shown in the last row of (\ref{eq.Lagrange1}). This implies that the SM Yukawa
couplings arise only through the coupling of the Higgs to the composite fermions and their
mixings to the elementary fermions. We further assume that the strong sector is flavor
anarchic, so that the hierarchy in the masses and mixings of the SM quarks comes from
the hierarchy in the mixing parameters $\Delta^i_{L,R}$. In this case the mixing
parameters of the light elementary quarks can be safely neglected and one can focus on
just the third generation of composite fermions. 
\footnote{In fact, once produced, heavy fermions of the first two generations will also decay mostly to tops and
bottoms, since flavor-changing transitions are not suppressed in the strong sector, while the couplings to the
light SM quarks are extremely small, see the discussion in Ref. \cite{Sundrum}.
}\\
As a consequence of the elementary/composite mass mixings, 
the top and the bottom masses arise, after the EWSB, from the Yukawa terms in the Lagrangian (\ref{eq.Lagrange1}), 
$Y_{*U}Tr\left\{\bar{\mathcal{Q}}\mathcal{H}\right\}\tilde{T}$ and $Y_{*D}Tr\left\{\bar{\mathcal{Q'}}\mathcal{H}\right\}\tilde{B}$. 
The top mass will be proportional to $\Delta_{L1}\Delta_{R1}$ and the bottom mass to $\Delta_{L2}\Delta_{R2}$. 
The small ratio between the bottom and the top quark masses can be thus obtained both for $\Delta_{L2}\ll\Delta_{L1}$ ($\Delta_{R2}\sim\Delta_{R1}$) 
and for $\Delta_{R2}\ll\Delta_{R1}$ ($\Delta_{L2}\sim\Delta_{L1}$). \\
For $t_{R}$, $b_{R}$ and their excited states the rotation from the elementary/composite basis to the mass eigenstate one, the SM/heavy basis, 
is given by:
\begin{align}
\begin{split}\label{rotation_RR2}
&\tan\varphi_{R}=\frac{\Delta_{R1}}{M_{\tilde{T}*}}\ \ s_{R}\equiv\sin\varphi_{R} \ \ c_{R}\equiv\cos\varphi_{R}  \\ 
&\tan\varphi_{bR}=\frac{\Delta_{R2}}{M_{\tilde{B}*}}\ \ s_{bR}\equiv\sin\varphi_{bR} \ \ c_{bR}\equiv\cos\varphi_{bR} \\  
&\left\{\begin{array}{l}
t_{R}=c_{R}t^{el}_{R}-s_{R}\tilde{T}^{com}_{R}\\
\tilde{T}_{R}=s_{R}t^{el}_{R}+c_{R}\tilde{T}^{com}_{R} \end{array} \right.\ 
\left\{\begin{array}{l}
		b_{R}=c_{bR}b^{el}_{R}-s_{bR}\tilde{B}^{com}_{R}\\ 
	\tilde{B}_{R}=s_{bR}b^{el}_{R}+c_{bR}\tilde{B}^{com}_{R} \end{array}  \right.
\end{split}
\end{align}
\noindent
$s_R(s_{bR})$ defines the degree of compositeness, $\xi_{tR}(\xi_{bR})$, of $t_R(b_R)$; $c_R(c_{bR})$ that of $\tilde{T}(\tilde{B})$, $\xi_{\tilde{D}}$. 
We will diagonalize analytically the mixing among $q^3_{L}$ and the corresponding excited states by requiring the simplifying assumption:
$\Delta_{L2}\ll\Delta_{L1}$, 
that can naturally follow, for example, from the RG flow in the full theory \cite{mchm2}. 
The first two generations of elementary quarks do not need a field rotation 
from the elementary/composite basis to the mass eigenstate basis, 
since they do not mix with the composite fermions and can thus be directly identified with the corresponding SM states.\\
We can see that in this model $t_{R}$ and $b_{R}$ are both $P_{C}$ and $P_{LR}$ eigenstates, 
since they couple to $SU(2)_{L}\times SU(2)_{R}$ singlets 
($T_L (\tilde{T},\tilde{B})=T_R (\tilde{T},\tilde{B})$, $T^3_{L} (\tilde{T},\tilde{B})=T^3_{R} (\tilde{T},\tilde{B})=0$).
Instead, $t_L$ is a $P_{LR}$ eigenstate only in the limit ($\Delta_{L1}=0$) in which 
it decouples from $T$ ($T^3_{L}(T)\neq T^3_{R}(T)$). Similarly, 
$b_L$ is a $P_{LR}$ eigenstate only for $\Delta_{L2}= 0$, in which case it decouples from $B'$ ($T^3_{L}(B')\neq T^3_{R}(B')$).\\
\noindent
So far we have made field rotations to the mass eigenstate basis before the EWSB. 
After the EWSB, the SM top and bottom quarks acquire a mass, and the heavy masses get corrections 
of order $\left(\frac{Y_{*}v}{\sqrt{2}m_{*}}\right)^2$. 
In the following, we assume $x \equiv \left(\frac{Y_{*}v}{\sqrt{2}m_{*}}\right) \ll 1$ and compute all quantities at leading order in 
$x$.

\subsubsection{$\Delta_{L2}\ll\Delta_{L1}$} 
In this case, since $\Delta_{L2}\ll\Delta_{L1}$, $b_{L}$ is, approximately, a $P_{LR}$ eigenstate so, approximately, we have a 
custodial symmetry protection to $Zb_{L}\bar{b}_L$.\\
The small ratio between the bottom and the top quark masses is obtained for $\Delta_{L2}\ll\Delta_{L1}$ ($\Delta_{R2}\sim\Delta_{R1}$); we have:
\begin{equation}
	m_{t}=\frac{v}{\sqrt{2}}Y_{*U}s_{1}s_{R}
\label{tmass}
\end{equation}
\begin{equation}
	m_{b}=\frac{v}{\sqrt{2}}Y_{*D}s_{2}s_{bR} \ ,
\label{bmass}
\end{equation}
\noindent
where $s_{1}=\sin\varphi_{L1}=\frac{\Delta_{L1}}{\sqrt{M^{2}_{Q*}+\Delta^{2}_{L1}}}$ defines the $(t_L,b_L)$ degree of compositeness, $\xi_{qL}$, and $s_{2}$ is a rotation angle proportional to $\Delta_{L2}$, 
$s_{2}=\frac{\Delta_{L2}}{M_{Q'*}}\cos\varphi_{L1}$. \\
\noindent 
The physical masses of the heavy fermions read:
\begin{align}
\left\{\begin{array}{l}
M_{\tilde{T}}=\sqrt{M^{2}_{\tilde{T}*}+\Delta^{2}_{R1}}\\
M_{\tilde{B}}=\sqrt{M^{2}_{\tilde{B}*}+\Delta^{2}_{R2}}\\
M_{T}=M_{B}=\sqrt{M^{2}_{Q*}+\Delta^{2}_{L1}} \\
M_{T5/3}=M_{T2/3}=M_{Q*}=M_T c_1 \\
M_{T'}=M_{B'}=\sqrt{M^{2}_{Q'*}+\Delta^{2}_{L2}}\simeq M_{Q'*}\\
M_{B-1/3}=M_{B-4/3}=M_{Q'*} \end{array} \right.
\label{HfMass_ts5}
\end{align}
where $c_1\equiv \cos\varphi_{L1}$ is the degree of compositeness, $\xi_{D}$, of the $SU(2)_L$ doublet $D=(T,B)$. Details can be found in App. \ref{TS5A}.\\
In order for the strong sector to respect the custodial invariance, as we have shown,
 composite fermions have to fill multiplets of $SU(2)_{L}\times SU(2)_{R}\times U(1)_{X}$. 
As a consequence, the heavy partner of the SM doublet $q^3_L=(t_L,b_L)$, $D=(T,B)$ ($=2_{1/6}$ under the SM electroweak group), is embedded in a larger multiplet, 
the bidoublet $\mathcal{Q}_{2/3}=(2,2)_{2/3}$, that includes an other doublet of heavy fermions, $(T_{5/3},T_{2/3})$($=2_{7/6}$). 
The heavy fermions $T_{5/3}$ and $T_{2/3}$ in this latter doublet are called \textit{custodians}. 
They share the same multiplet of the heavy partners of $q^3_L$ but they do not mix directly with the SM fermions. 
This implies that their masses tend to zero in the limit in which $t_L$ becomes fully composite (see for example the discussion in \cite{pomarol_serra}). 
This can be seen from eq. (\ref{HfMass_ts5}): $M_{T5/3(2/3)}$ is zero for $c_1=0$, i.e. for a fully composite $t_L$ ($s_1=1$).


\subsection{TS10}\label{TS10}

In TS10 we consider composite fermions embedded into a $\mathbf{10_{2/3}}$ representation 
of $SO(5)\times U(1)_{X}$, that decomposes as $\mathbf{10_{2/3}}=\mathbf{(2,2)_{2/3}}\oplus\mathbf{(1,3)_{2/3}}\oplus\mathbf{(3,1)_{2/3}}$ under 
$SU(2)_{L}\times SU(2)_{R}\times U(1)_{X}$. Therefore we refer to this field content in the composite sector:
\begin{eqnarray}	
\nonumber
	\mathcal{Q}_{2/3}=\left[\begin{array}{cc}
	T & T_{5/3} \\ 
	B & T_{2/3} \end{array}\right]=\left(2,2\right)_{2/3}\\ \nonumber 
 \mathcal{\tilde{Q}}_{2/3}=\left(\begin{array}{c}
	\tilde{T}_{5/3} \\ 
	\tilde{T}\\
        \tilde{B} \end{array}\right)=\left(1,3\right)_{2/3}\ , \  \mathcal{\tilde{Q}'}_{2/3}=\left(\begin{array}{c}
	\tilde{T}'_{5/3} \\ 
	\tilde{T}' \\
        \tilde{B}' \end{array}\right)=\left(3,1\right)_{2/3}   \\ \nonumber
	\mathcal{H}=\left[\begin{array}{cc}
	\phi^{\dag}_{0} & \phi^{+} \\
	-\phi^{-} & \phi_{0} \end{array}\right]=\left(2,2\right)_{0} \\
	\label{eq.fields_mchm10}
\end{eqnarray}
and to the following fermionic Lagrangian in the elementary/composite basis:

\begin{align}
\begin{split}\label{eq.Lagrange1_mchm10}
\mathcal{L}= \ &\bar{q}^3_{L} i\dslash q^3_{L}+\bar{t}_{R}i\dslash t_{R}+\bar{b}_{R}i\dslash b_{R} \\
& +Tr\left\{\bar{\mathcal{Q}}\left(i\dslash -M_{Q*}\right)\mathcal{Q}\right\}+Tr\left\{\mathcal{\bar{\tilde{Q}}}\left(i\dslash-M_{\tilde{Q}*}\right)\mathcal{\tilde{Q}}\right\}
+Tr\left\{\mathcal{\bar{\tilde{Q}}'}\left(i\dslash-M_{\tilde{Q}*}\right)\mathcal{\tilde{Q}}'\right\}\\ 
& +Y_{*}Tr\left\{\mathcal{H}\bar{\mathcal{Q}}\mathcal{\tilde{Q}'}\right\}+Y_{*}Tr\left\{\bar{\mathcal{Q}}\mathcal{H}\mathcal{\tilde{Q}}\right\} \\ 
& -\Delta_{L1}\bar{q}^3_{L}\left(T,B\right)-\Delta_{R1}\bar{t}_{R}\tilde{T}-\Delta_{R2}\bar{b}_{R}\tilde{B}+h.c.\ .
\end{split}	
\end{align}
\noindent
We have the following expressions for the top and bottom masses:
\begin{equation}
 m_{t}=\frac{v}{2}Y_{*}s_{1}s_{R} \ , \ \  m_{b}=\frac{v}{\sqrt{2}}Y_{*}s_{1}s_{bR}
\end{equation} 
and for the heavy fermion physical masses:
\begin{align}
\left\{\begin{array}{l}
M_{\tilde{T}}=\sqrt{M^{2}_{\tilde{Q}*}+\Delta^{2}_{R1}}\\
M_{\tilde{B}}=\sqrt{M^{2}_{\tilde{Q}*}+\Delta^{2}_{R2}}=M_{\tilde{T}} c_R / c_{bR}\simeq M_{\tilde{T}} c_R\\
M_{\tilde{T}5/3}=M_{\tilde{T}'5/3}=M_{\tilde{T}'}=M_{\tilde{B}'}=M_{\tilde{T}} c_R\\
M_{T}=M_B=\sqrt{M^{2}_{Q*}+\Delta^{2}_{L1}} \\
M_{T2/3}=M_{T5/3}= M_T c_1 \end{array} \right. \ .
\end{align}
\noindent
More details can be found in App. \ref{TS10A}.\\
Besides the custodians $T_{5/3}$ and $T_{2/3}$, which are light in the case of a composite $q^3_L$, $\tilde{T}_{5/3}$ and the fermions
 in the $\mathcal{\tilde{Q}'}_{2/3}$ triplet become light for a $t_R$ with a large degree of compositeness (also $\tilde{B}$ becomes light in this case).\\
In this model, both $t_{R}$ and $b_{R}$ are not $P_{LR}$ eigenstates and only $t_R$ is a $P_{C}$ eigenstate, 
as a consequence of the couplings to $\mathcal{\tilde{Q}}$ 
($T_L(\tilde{T},\tilde{B})\neq T_R(\tilde{T},\tilde{B})$; 
in particular, $b_R$ is not a $P_C$ eigenstate, since $T^3_{R}(\tilde{B})\neq 0$. $b_{L}$ is exactly a $P_{LR}$ eigenstate.


\subsection{$Zb_{L}\bar{b}_{L}$ in the TS Models}
Shifts in the $Z$ coupling to $b_{L}$, $g_{Lb}$, have been extensively studied in the literature. See, for example, the studies \cite{Zbb-RS} in the context of Randall-Sundrum models and \cite{Zbb-TS} in two-site descriptions. The shifts arise after the EWSB because of electroweak mixings among $b_{L}$ and heavy fermions. There is also a contribution 
from the mixing among neutral gauge bosons; however this mixing is of the order $(\frac{v}{M_{*}})^{2}\ll 1$, 
where $M_{*}$ stands for the heavy
neutral boson mass, and we will neglect it in what follows.\\
In two-site models without $P_{LR}$ symmetry 
there is no custodial symmetry protection to $Zb_{L}\bar{b}_{L}$ and so the shift on $g_{Lb}$ is large. 
Naive Dimensional Analysis (NDA) \cite{NDA} gives (see, for example, \cite{Agashe_flav, SILH}):
\begin{equation}
 \frac{\delta g_{Lb}}{g_{Lb}}\sim \frac{m^{2}_{t}}{M^{2}_{Q*}s^{2}_{R}}\sim \frac{Y^2_{*} v^2 s^2_1}{M^2_{Q*}}\ .
\label{gLb_TS}
\end{equation} 
\noindent
This formula has been obtained by approximating $q^2=M^2_Z\simeq 0$. 
At $q^2=M^2_Z$ the shift receives $O\left(\frac{M^2_Z}{M^2_{Q*}}\right)$ corrections:
\begin{equation}
 \frac{\delta g_{Lb}}{g_{Lb}}\sim \frac{M^2_Z s^2_1}{M^2_{Q*}}\sim \left( \frac{v^2 Y^2_{*} s^2_1}{M^2_{Q*}}\right)\frac{g^2}{Y^2_{*}}\ . 
\label{gLb_QnonZero}
\end{equation}
When compared to (\ref{gLb_TS}), there is a suppression $\left( \frac{g}{Y_{*}}\right)^2$ (see for example \cite{contino_fcnc}), so we will neglect it in the following.

LEP and SLD experiments fix an upper bound of $0.25\%$ for the (positive) shift in the $g_{Lb}$ from its SM value. Therefore, from
the eq. (\ref{gLb_TS}), we derive the following bound for the heavy fermion mass in models without custodial symmetry protection to $Zb_{L}\bar{b}_{L}$:
\begin{equation}
M_{Q*}\gtrsim (3.2)\frac{1}{s_{R}} \ \text{TeV} \ .    
\label{no_cust_bound}                                    
 \end{equation} 
\noindent
In order to respect this limit without requiring too large heavy fermion masses, that would contrast with naturalness arguments, it is necessary
to have a quite composite right-handed top (i.e., a not small $s_{R}$). On the contrary, in models with custodial symmetry protection to $Zb_{L}\bar{b}_{L}$, 
there is no such restriction for the $t_{R}$ degree of compositeness and bounds are weaker than the one in (\ref{no_cust_bound}). Indeed, in the TS5
with $\Delta_{L2}\ll\Delta_{L1}$, where we have approximately a custodial symmetry protection to $Zb_{L}\bar{b}_L$ 
(the breaking is proportional to $\Delta_{L2}$ and is thus small), we obtain:
\begin{align}
\begin{split} 
 \frac{\delta g_{Lb}}{g_{Lb}} =\left(\frac{Y_{*}v}{\sqrt{2}} \right)^{2}\left(\frac{s_{2}c_{bR}}{\sqrt{2}M_{\tilde{B}}}\right)^{2}[T^3_{L}(\tilde{B})-T^3_{L}(b_{L})]
=\frac{1}{2}\frac{m^{2}_{b}}{M^{2}_{Q*}}\frac{c^{4}_{bR}}{s^{2}_{bR}}\simeq\frac{1}{2}\frac{m^{2}_{t}}{M^{2}_{Q*}}\frac{s^{2}_{2}}{s^{2}_{R}} \ .
\end{split}
\end{align}
As expected, the shift is proportional to $s^{2}_{2}$ (i.e., it is proportional to $\Delta^{2}_{L2}$, the size of the
custodial symmetry breaking) and it is small (notice that is also smaller than the effect at non-zero momentum).
In the TS10, we obtain, again, a small shift:
\begin{align}
 \begin{split}
  \frac{\delta g_{Lb}}{g_{Lb}}= & \left(\frac{Y_{*}v}{\sqrt{2}}\right)^{2}\frac{s^{2}_{1}}{M^{2}_{Q*}}\left[c^{4}_{bR}(T^3_{L}(\tilde{B})-T^3_{L}(b_{L}))+(T^3_{L}(B')-T^3_{L}(b_{L}))\right]\\
  = & -\frac{m^{2}_{b}}{M^{2}_{Q*}}\frac{2-s^{2}_{bR}}{2}\simeq -\frac{m^{2}_{b}}{M^{2}_{Q*}}\ .
 \end{split}
\end{align}
Despite $b_L$ is an exact $P_{LR}$ eigenstate in the TS10, there is still a small modification that comes from the coupling of $b_{R}$, 
that explicitly breaks $P_{LR}$. Notice that $\delta g_{Lb}=0$, if we have $s_{bR}=0$.

\section{Bounds from flavor observables}\label{sec:bound}
\subsection{Constraint from the process $b\rightarrow s\gamma$}\label{bsGamma}

\begin{figure}
\centering
\scalebox{0.7}{\includegraphics{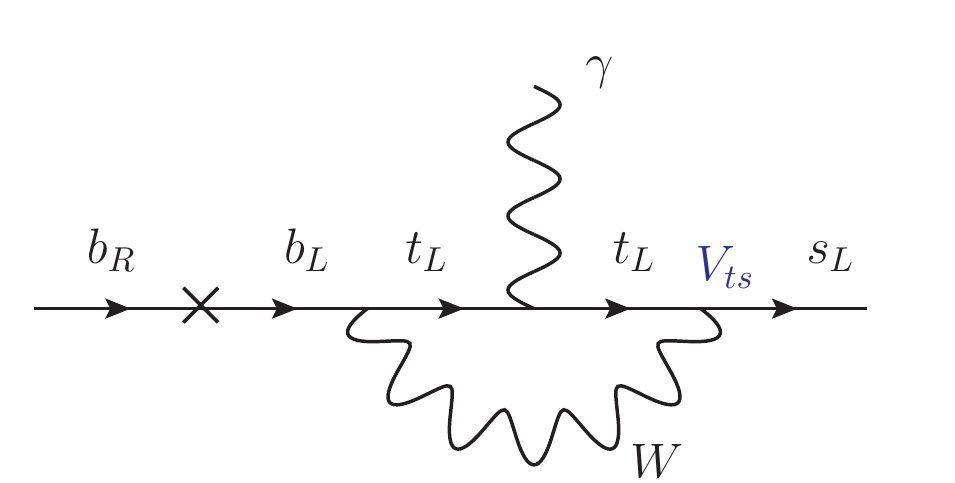}}
\caption{\small{1 loop Infrared contribution to $C_7$ in the SM.}}
\label{C7-sm}
\end{figure}

We define, following \cite{Buras}, the effective Hamiltonian for $b\rightarrow s\gamma$:
\begin{equation}
 \mathcal{H}_{eff}=-\frac{G_{F}}{\sqrt{2}}V^{*}_{ts}V_{tb}\left[C_{7}(\mu_{b})\mathcal{O}_{7} + C^{'}_{7}(\mu_{b})\mathcal{O}^{'}_{7}\right] \ ,
\label{Heff} 
\end{equation} 
where $\mathcal{O}_{7}=\frac{e}{8\pi^{2}}m_{b}\bar{b}\sigma^{\mu\nu}F_{\mu\nu}(1-\gamma_{5})s$ and 
$\mathcal{O}^{'}_{7}=\frac{e}{8\pi^{2}}m_{b}\bar{b}\sigma^{\mu\nu}F_{\mu\nu}(1+\gamma_{5})s$.\\
In the SM the $W$ boson has a purely $V-A$ interaction to the fermions and so the contribution to the $b\rightarrow s\gamma$
process has to proceed through mass insertions in the external legs (see Fig. \ref{C7-sm}). The Wilson coefficient $C'_{7}$ is thus negligible, 
because of a suppression by a factor $m_{s}/m_{b}$ in respect to the Wilson coefficient $C_{7}$,
that, evaluated at the weak scale $\mu_{w}$ is \cite{Buras}
  \begin{equation}
   C^{SM}_{7}(\mu_{w})=-\frac{1}{2}\left[ -\frac{(8x^{3}_{t}+5x^{2}_{t}-7x_{t})}{12(1-x_{t})^{3}}+\frac{x^{2}_{t}(2-3x_{t})}{2(1-x_{t})^{4}}\ln(x_{t})\right] \ ,
  \end{equation} 
with $x_{t}=\frac{m^{2}_{t}}{M^{2}_{W}}$. \\
In composite Higgs models there are two classes of effects that lead to a shift of the $b\rightarrow s\gamma$ decaying rate
compared to the Standard Model prediction. 
The first comes from loops of heavy fermion resonances from the strong sector that generate the flavor-violating dipole operators 
$\mathcal{O}_{7}$, $\mathcal{O}^{'}_{7}$ at the compositeness scale. 
We will refer to this as the UV contribution. The second contribution comes from the tree level exchange of heavy resonances, 
which generates an effective V+A interaction of the $W $boson and the SM quarks which in turn leads to
a shift to $b\to s\gamma$ via a loop of SM particles. This latter IR contribution is enhanced by a chiral factor $m_t/m_b$. Since in
this case the flavor violation can come entirely from the SM V-A current, it gives a quite
model-independent lower bound on the heavy fermion masses.\\
By taking into account the experimental average value for the $b\rightarrow s\gamma$ branching ratio \cite{BsGamma_exp} and the 
theoretical calculation \cite{BsGamma_th}, we get, if the new physics contributions to $C_{7}$, $C^{CH}_{7}$, and to $C^{'}_{7}$, $C^{'CH}_{7}$, are considered
separately, the bounds (see Appendix \ref{App_bound}):
\begin{equation}
 -0.098 \lesssim C^{CH}_{7}(m_{*}) \lesssim 0.028 
\label{C7UVlim}
\end{equation}  
\begin{equation}
 |C^{'CH}_{7}(m_{*})| \lesssim 0.37 \ ,
\label{C'7lim}
\end{equation}  
where $m_{*}$ denotes the mass of the heavy fermions in the loop (we take $m_{*}=1$ TeV). \\
The infrared contribution to $b\rightarrow s\gamma$ from the composite Higgs model is at the weak scale $\mu_{w}$ instead of $m_{*}$ (we take $\mu_W = M_W$); 
therefore, we have to account for a scaling factor
\begin{equation}
 C^{CH}_{7}(\mu_{w})=\left[ \frac{\alpha_{s}(m_{*})}{\alpha_{s}(m_{t})}\right]^{16/21} \left[\frac{\alpha_{s}(m_{t})}{\alpha_{s}(\mu_{w})} \right]^{16/23} C^{CH}_{7}(m_{*}) \approx 0.79  C^{CH}_{7}(m_{*})
\end{equation}  
We get:
\begin{equation}
 -0.077\lesssim C^{CH}_{7}(\mu_{w}) \lesssim 0.023
\label{c7limIR}
\end{equation}  
\begin{equation}
 |C^{'CH}_{7}(\mu_{w})| \lesssim 0.29
\label{C'7limIR}
\end{equation}  
\noindent
While the infrared contribution to $C_7$ involves a flavor-conserving operator and brings to a MFV bound, 
the infrared contribution to $C^{'}_7$ as well as the ultraviolet contributions to $C_7$ and to $C^{'}_7$ involve 
flavor-violating operators. As a consequence, they will require some assumptions on the flavor structure of the NP sector.    

\noindent
We will now evaluate the bounds on heavy masses that come from the infrared contribution to $C_7$. 
We will first present estimates of such bounds in generic composite Higgs models, which can be obtained by NDA. 
Then we will calculate the bounds in the specific two-site model TS5 and TS10, introduced in sec.s \ref{TS5} and \ref{TS10}.

\subsection{MFV bound from the infrared contribution to $C_7$}\label{IR_contr}

\begin{figure}
\centering
\scalebox{0.7}{\includegraphics{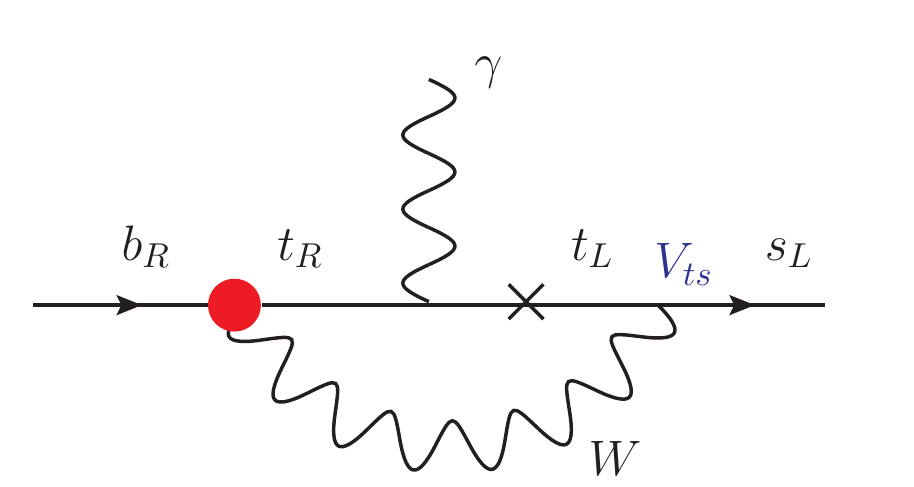}}
\caption{\small{1 loop Infrared contribution to $C_7$. 
The red blob denotes the effective coupling $Wt_R b_R$, generated from the composite sector.}}
\label{fig:C7IR_dia}
\end{figure}

The infrared contribution to the process $b\rightarrow s\gamma$ is a one loop contribution from
the $W$ boson accompanied by top quarks, where a mass insertion in the intermediate top quark states is allowed by the presence of a 
$(V+A)$ interaction of the $W$ boson with the top and the bottom quarks (Fig. \ref{fig:C7IR_dia}).
This interaction originates from a term:
\begin{equation}
\mathcal{L}\supset \mathcal{C}_{R}\mathcal{O}_{R}\ ,
\label{wright_term}
\end{equation} 
where $\mathcal{O}_{R}$ is the dimension-6 operator:
\begin{equation}
\mathcal{O}_{R}\equiv H^{c\dag}iD_{\mu}H\bar{t}_{R}\gamma^{\mu}b_{R}+h.c.\ .
\label{oWR}
\end{equation}  
At low energy, after the EWSB, the interaction in (\ref{wright_term}) gives:
\begin{equation}
\mathcal{L}\supset \frac{\mathcal{C}_{R}v^{2}}{2}\frac{g_{2}}{\sqrt{2}}\bar{b}_{R}\gamma^{\mu}t_{R}W^{-}_{\mu}\ .
\end{equation} 
\noindent
This interaction gives a contribution to the Wilson coefficient $C_{7}$ in the eq. (\ref{Heff}). We find:
\begin{equation}
C^{CH-IR}_{7}(\mu_{w})= \frac{\mathcal{C}_{R}v^{2}}{2}\frac{m_{t}}{m_{b}}f_{RH}(x_{t})
\label{CchIR}
\end{equation} 
where $x_{t}=\frac{m^{2}_{t}}{M^{2}_{W}}$ and $f_{RH}(x_{t})$ is the loop function \cite{ARH}:

\begin{align}
\begin{split}
 f_{RH}(x_{t})=\ & -\frac{1}{2}\left\{ \frac{1}{\left(1-x_{t}\right)^{3}}\frac{2}{3}\left[ -\frac{x^{3}_{t}}{2}-\frac{3}{2}x_{t}+2+3x_{t}\log(x_{t})\right]\right.\\
& + \frac{1}{\left(1-x_{t}\right)^{3}} \left[ -\frac{x^{3}_{t}}{2}+6x^{2}_{t}-\frac{15}{2}x_{t}+2-3x^{2}_{t}\log(x_{t})\right] \Bigg\} \ .
 \end{split}
\end{align}
\noindent
$f_{RH}=-0.777$, for $m_t=174$ GeV and $M_W=80.4$ GeV.\\

\begin{figure}
\centering
\scalebox{0.7}{\includegraphics{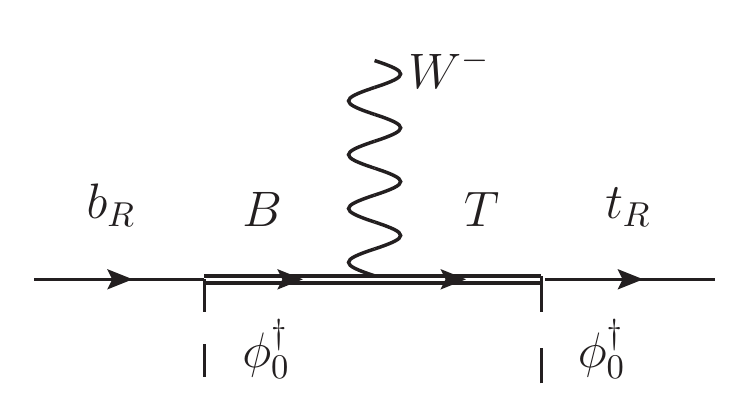}}
\caption{\small{The CHM contribution to the effective coupling $Wt_R b_R$ (At order $\left( \frac{Y_* v}{\sqrt{2}m_*}\right)^2 $).}}
\label{wtRbR_diagram}
\end{figure}

\noindent
We point out that the bound on the CHM contributions to $b \to s\gamma$, $C^{CH}_{7}$ in eq. (\ref{c7limIR}), 
can be directly translated into a bound on the effective vertex $Wt_R b_R$, $v_R\equiv\frac{\mathcal{C}_{R}v^{2}}{2}$. 
By considering the bound in (\ref{c7limIR}) and the relation in (\ref{CchIR}), we obtain:
\begin{equation}
-0.0004<v_R<0.0013\ .
\label{bound_vR}
\end{equation} 

This bound from $b\to s \gamma$ can be compared with that from the measurement of the $Wtb$ anomalous couplings at colliders. 
Ref. \cite{saavedra_vR} reports an expected bound of $-0.012<v_R<0.024$, that can be imposed by 14 TeV LHC measurements with $30$ fb$^{-1}$. 
This latter can be obtained from studies on cross sections and top decay observables (angular distributions and asymmetries) 
in the single top production at the LHC. 
Present searches for anomalous $W$ couplings at the 7 TeV LHC \cite{LHC_vR} fix still mild bounds on $v_R$, $-0.34<v_R<0.39$, with $0.70$ fb$^{-1}$.
We can see that the bound obtained from $b\to s \gamma$ is much stronger than that from the $v_R$ measurement at collider.\\

\noindent
The CHM contribution to the effective coupling $Wt_R b_R$ is given 
by the exchange of heavy fermions that mix electro-weakly with $t_R$ and $b_R$ (fig. \ref{wtRbR_diagram}). 
At the order $x^2 $, only the $SU(2)_L$ heavy doublets which are partners of $(t_L,b_L)$ contribute to $\mathcal{C}_{R}$. 
This latter can be easily estimated by NDA \cite{NDA}:
\begin{equation}
 \mathcal{C}_{R}\sim \frac{ Y^2_{*} \xi_{bR}\xi_{tR}\xi^2_D}{M^2_D}\sim\frac{y_{b}y_{t}}{M^{2}_{D}}\frac{\xi^2_D}{\xi^2_{qL}} \ .
\label{CR_est}
\end{equation} 
(\ref{CR_est}) implies:
\begin{equation}
C^{CH-IR}_{7}(\mu_{w})\sim\frac{m^{2}_{t}}{M^{2}_{D}}f_{RH}(x_{t}) \frac{\xi^2_D}{\xi^2_{qL}}\ .
\label{CchIR_est}
\end{equation}
Applying the condition in (\ref{c7limIR}) to this infrared contribution, we get the estimated bound:
\begin{equation}
\mathbf{M_D \gtrsim \frac{1.0 (0.54)}{\xi_{qL}}}\ \textbf{TeV} \ ,
\label{C7bound_est}
\end{equation}
where the first number and the second number in parenthesis refer respectively to the case 
of a positive and of a negative $C^{CH-IR}_{7}$ contribution. 
Notice that in the case of a positive $C^{CH-IR}_{7}$ contribution we obtain a stronger bound on $M_D$, 
since the constraint in (\ref{c7limIR}) is asymmetric.\\
\noindent
We find that a subgroup of the custodial symmetry $SU(2)_V$, the $P_C$ parity, can give a suppression to the $Wt_R b_R$ coupling
and, as a consequence, to the CHM infrared contribution to $b\to s \gamma$.
The estimates we have just reported refer to generic composite Higgs models where there is not such $P_C$ protection.

\subsubsection{Protection by $P_C$ parity}\label{PCprotect}
The $P_C$ protection against the generation of the $Wt_R b_R$ vertex acts similarly to the $P_{LR}$ and $P_C$ protection against large corrections
to the $Zb_L b_L$ coupling, which we have discussed in sec. \ref{PLR}. 
$P_C$ is a symmetry of the sector BSM, that is respected also by the interactions of $t_R$ and $b_R$ if these latter are $P_C$ eigenstates. 
Since $P_C$ acts as $diag(1,-1,-1)$ on $SO(3)$ vectors, the $W$ is not a $P_C$ eigenstate (the composite partners of $W^1$ and $W^2$ have not the same $P_C$ eigenvalue). 
In the case in which $t_R$ and $b_R$ are both $P_C$ eigenstates,
both the $t_R$ and the $b_R$ interactions must respect the $P_C$ parity. Then, the $Wt_R b_R$ vertex, which is $P_C$ violating,
since the $W$ is not a $P_C$ eigenstate, can arise only by paying for an additional factor, that gives a suppression. 
Whereas, in models where $t_R$ and $b_R$ are not both $P_C$ eigenstates and, as such, 
their interactions have not to respect the $P_C$ parity, the $Wt_R b_R$ vertex can be generated without suppressions.\\
 The TS5 falls into the class of models with $P_C$ protection, since in the TS5 both $t_R$ and $b_R$ are $P_C$ eigenstates. 
Considering the TS5, we can evaluate the suppression factor to $Wt_R b_R$ due to the $P_C$ protection. 
We can find it in an easy way by promoting $\Delta_{L1}$ and $\Delta_{L2}$ to spurions, 
which enforce a $SU(2)_L\times SU(2)_R$ invariance:
\[
 -\Delta_{L1}\bar{q}^3_L \left( T, B\right) \to \ -\bar{q}^3_L \mathcal{Q}_{2/3}\hat{\Delta}_{L1}
\]
\[
 -\Delta_{L2}\bar{q}^3_L \left( T', B'\right) \to \ -\bar{q}^3_L \mathcal{Q'}_{-1/3}\hat{\Delta}_{L2} \ ,
\]
where $\hat{\Delta}_{L1}=(\Delta_{L1},0)\equiv (1,2)_{1/6}$ and $\hat{\Delta}_{L2}=(0,\Delta_{L2})\equiv (1,2)_{1/6}$.
We can thus write the $\mathcal{O}_R$ operator (\ref{oWR}) in the $SU(2)_L\times SU(2)_R$ invariant way:
\begin{equation}
 \mathcal{O}_R = \frac{1}{f^2}\bar{q}^3_R \hat{\Delta}_{L1}V_{\mu}\hat{\Delta}^{\dag}_{L2}q^3_R \gamma^{\mu} + h.c. \ ,
\label{OR_inv}
\end{equation}
where $f$ has the dimension of a mass, $q^3_R=(t_R, b_R)\equiv (1,2)_{1/6}$ 
and $V_{\mu}\equiv H^{c\dag}iD_{\mu}H$. 
Since $P_C$ is a subgroup of the custodial $SU(2)_V$, the $SU(2)\times SU(2)$ invariant operator in (\ref{OR_inv}) is also a $P_C$ invariant. 
We can notice that the $P_C$ invariance has brought to an additional factor $\frac{\Delta_{L1}\Delta_{L2}}{f^2}$ compared to (\ref{oWR}). \\
Without $P_C$ protection, the $D= (T,B)$ contribution to the $Wt_R b_R$ effective vertex in the TS5 reads
\[
 s_R s_{bR} c^2_1 \left(\frac{Y_* v}{\sqrt{2} M_D} \right)^2=\frac{m_b m_t}{M^2_D}\frac{c^2_1}{s^2_1}\ ;
\]
the request for $P_C$ invariance brings to the additional factor $\frac{\Delta_{L1}\Delta_{L2}}{f^2}$. For $f^2=M_{Q*}M_{Q'*}$, we obtain
\[
  \left(\frac{Y_* v}{\sqrt{2} M_D} \right)^2 s_R s_{bR}\frac{c_1\Delta_{L1}}{M_{Q*}}\frac{c_1\Delta_{L2}}{M_{Q'*}}=\left(\frac{Y_* v}{\sqrt{2} M_D} \right)^2 s_R s_{bR}s_1 s_2=\frac{m_b m_t}{M^2_D}\ ,
\]
  that is a suppression by a factor $s^2_1/c^2_1\equiv \xi^2_{qL}/\xi^2_D$. \\ 
\noindent

We can thus return to the estimated bounds on $M_D$ from $C^{CH-IR}_7$ in eq. (\ref{C7bound_est}), and consider the case in which there is a $P_C$ protection 
to the $t_R$ and $b_R$ interactions. In such case the $\mathcal{C}_{R}$ contribution becomes:

\begin{equation}
 \mathcal{C}_{R} \sim\frac{y_{b}y_{t}}{M^{2}_{D}}  \ \ (\text{with}\ P_C) \ ,
\label{CR_est_pc}
\end{equation} 
which implies
\begin{equation}
C^{CH-IR}_{7}(\mu_{w})\sim\frac{m^{2}_{t}}{M^{2}_{D}}f_{RH}(x_{t}) \ \ (\text{with}\ P_C) \ 
\label{CchIR_est_pc}
\end{equation}
and an estimated bound:
\begin{equation}
\mathbf{M_D \gtrsim 1.0 (0.54)}\ \textbf{TeV} \ (\text{with}\ P_C) \ .
\label{C7bound_est_pc}
\end{equation}

We will now calculate the bounds on $M_D$ from $C^{CH-IR}_7$ in the specific TS5 and TS10 models. 
As already discussed, the TS5 belongs to the class of models with $P_C$ protection. 
The TS10, instead, falls in the class of models without $P_C$ protection, because in the TS10 $b_R$ is not a $P_C$ eigenstate. 
We thus expect that the bound in the TS10 will receive an enhancement factor $c_1/s_1$, compared to that in the TS5.\\
 
\noindent
In the TS5 we have a contribution to the $\mathcal{O}_R$ operator in (\ref{oWR}) both from the doublet $D= (T,B)$ in the $X=2/3$ representation and from the doublet $D'\equiv (T',B')$ in the $X=-1/3$. 
We find: 
\begin{equation}
 \mathcal{C}^{TS5}_{R}=-\frac{y_{b}y_{t}}{M^{2}_{D}}\left( 1+\frac{M^{2}_{D}}{M^{2}_{D'}}\right) \ .
\end{equation} 
This implies:
\begin{equation}
C^{CH-IR-TS5}_{7}(\mu_{w})=-\frac{m^{2}_{t}}{M^{2}_{D}}f_{RH}(x_{t})\left( 1+\frac{M^{2}_{D}}{M^{2}_{D'}}\right) \ .
\label{CchIR-ts5}
\end{equation}
\noindent
Notice that the $\mathcal{C}^{TS5}_{R}$ contribution is negative. 
This implies a positive contribution $C^{CH-IR-TS5}_{7}$ ($f_{RH}$ is negative). 
The condition in (\ref{c7limIR}) is asymmetric and is stronger in the case of a positive $C^{CH-IR}_7$. 
Applying this condition to the infrared contribution in (\ref{CchIR-ts5}), we get, for $r=\frac{M_{D}}{M_{D'}}=1$, 
the following bound on the $D= (T,B)$ doublet mass:
\begin{equation}
\mathbf{ M^{TS5}_{D}\gtrsim 1.4}\ \textbf{TeV}\ .
\label{boundIR-TS5}
\end{equation} 
This bound becomes $M^{TS5}_{D}\gtrsim 1.3(1.6)$ TeV, changing $r$ to $r=0.8(1.2)$.
\noindent
In the TS10, there is only one doublet, $D=(T,B)$, that gives a contribution to $\mathcal{C}_{R}$. 
We obtain
\begin{equation}
 \mathcal{C}^{TS10}_{R}=\frac{y_{b}y_{t}}{M^{2}_{D}}\frac{c^{2}_{1}}{s^{2}_{1}}\ ,
\end{equation}
\noindent
which implies:
\begin{equation}
C^{CH-IR-TS10}_{7}(\mu_{w})=\frac{m^{2}_{t}}{M^{2}_{D}}f_{RH}(x_{t})\frac{c^{2}_{1}}{s^{2}_{1}} \ .
\label{CchIR-ts10}
\end{equation}
\noindent
From the condition in (\ref{c7limIR}) we get finally the bound:
\begin{equation}
\mathbf{ M^{TS10}_{D}\gtrsim (0.54)\frac{c_1}{s_1}}\ \textbf{TeV}\ .
\label{boundIR-TS10}
\end{equation} 
Notice that, differently from the case of the TS5 contribution, $C^{CH-IR-TS10}_{7}(\mu_{w})$ is negative. 
As such, it is constrained less strongly by the condition in (\ref{c7limIR}). 
As expected, we have found a $c_1/s_1$ enhancement of this bound, compared to (\ref{boundIR-TS5}).\\
\noindent
We now proceed to evaluate the bounds from the $C^{'}_7$ contribution and then those from the UV contributions. 
As we already pointed out, these are contributions that involve flavor-violating operators and 
require assumptions on the flavor structure of the NP sector.
In what follows we will consider the case of flavor anarchy of the composite Yukawa matrices. 
This scenario, we remind, assumes that there is no
large hierarchy between elements within each matrix $Y_*$ and the quark mass hierarchy is completely explained by the elementary/composite mixing angles. 
We also set, for simplicity, $Y_{*U}=Y_{*D}=Y_{*}$.

\begin{table}[]
\scalebox{0.9}{
\small{\begin{tabular}{|lccc}
\cline{1-3}
&\multicolumn{2}{|c|}{}&\\[0.015cm]
\multicolumn{1}{|l}{$C^{CH-IR}_{7}(\mu_w)$} &\multicolumn{2}{|c|}{$\sim\frac{\left(y_t v \right)^2 }{M^{2}_{D}}\xi^{2}_{D}  \qquad  w/\ P_C$}&\multicolumn{1}{c}{}\\
&\multicolumn{2}{|c|}{}&\\[0.01cm]
&\multicolumn{1}{|c}{ESTIMATED} &\multicolumn{1}{c|}{TS5}&\multicolumn{1}{c}{}\\
&\multicolumn{1}{|c}{$\mathbf{M_D \gtrsim 1.0 (0.54)}\ \textbf{TeV}$}&\multicolumn{1}{c|}{$\mathbf{M_D \gtrsim 1.4}\ \textbf{TeV}$}&\multicolumn{1}{c}{}\\
\textbf{MFV}&\multicolumn{2}{|c|}{}&\\[0.01cm]
\cline{2-3}
\textbf{Bounds} &\multicolumn{2}{|c|}{}&\\[0.015cm]
 &\multicolumn{2}{|c|}{$\sim\frac{\left(y_t v \right)^2 }{M^{2}_{D}}\left( \frac{\xi_D}{\xi_{qL}}\right)^2 \qquad w/o\ P_C $}&\multicolumn{1}{c}{}\\
 &\multicolumn{2}{|c|}{}&\\[0.01cm]
 &\multicolumn{1}{|c}{ESTIMATED} &\multicolumn{1}{c|}{TS10}&\multicolumn{1}{c}{}\\
 &\multicolumn{1}{|c}{$\mathbf{M_D \gtrsim 1.0 (0.54)/\xi_{qL}}$ \textbf{TeV}}&\multicolumn{1}{c|}{$\mathbf{M_D \gtrsim 0.54/s_{1}}$ \textbf{TeV}}&\\
&\multicolumn{2}{|c|}{}&\\[0.01cm]
\cline{1-3}
&\multicolumn{2}{|c|}{}&\\[0.015cm]
\multicolumn{1}{|l}{$C^{' CH-IR}_{7}(\mu_w)$} &\multicolumn{2}{|c|}{ $\sim\frac{\left(y_t v \right)^2 }{M^{2}_{D}}\xi^{2}_{D}\frac{m_s}{m_b V^{2}_{ts}} \qquad w/\ P_C $}&\multicolumn{1}{c}{}\\
&\multicolumn{2}{|c|}{}&\\[0.01cm]
&\multicolumn{1}{|c}{ESTIMATED} &\multicolumn{1}{c|}{TS5}&\multicolumn{1}{c}{}\\
&\multicolumn{1}{|c}{$M_D \gtrsim 0.80$ TeV}&\multicolumn{1}{c|}{$M_D \gtrsim 1.1\ \text{TeV}$}&\multicolumn{1}{c}{}\\
&\multicolumn{2}{|c|}{}&\\[0.01cm]
\cline{2-3}
&\multicolumn{2}{|c|}{}&\\[0.01cm]
 &\multicolumn{2}{|c|}{$\sim\frac{\left(y_t v \right)^2 }{M^{2}_{D}}\left( \frac{\xi_D}{\xi_{qL}}\right)^2 \frac{m_s}{m_b V^{2}_{ts}} \qquad w/o\ P_C $}&\\
 &\multicolumn{2}{|c|}{}&\\[0.01cm]
 &\multicolumn{1}{|c}{ESTIMATED} &\multicolumn{1}{c|}{TS10}&\multicolumn{1}{c}{}\\
 &\multicolumn{1}{|c}{$M_D \gtrsim 0.80/\xi_{qL}$ TeV}&\multicolumn{1}{c|}{$M_D \gtrsim 0.80/s_{1}$ TeV} &\\
 &\multicolumn{2}{|c|}{}&\\[0.01cm]
\cline{1-4}
&\multicolumn{3}{|c|}{}\\[0.015cm]
\multicolumn{1}{|l}{$C^{CH-UV}_{7}(m_{*})$} &\multicolumn{3}{|c|}{$\sim\frac{\left(Y_* v \right)^2 }{M_{D}M_{\tilde{D}}}\xi_{D}\xi_{\tilde{D}}$}\\
&\multicolumn{3}{|c|}{}\\[0.01cm]
&\multicolumn{1}{|c}{ESTIMATED} &\multicolumn{1}{c}{TS5}&\multicolumn{1}{c|}{TS10}\\
&\multicolumn{1}{|c}{$\sqrt{M_D M_{\tilde{D}}} \gtrsim 1.5 (0.79)Y_*$ TeV}&\multicolumn{1}{c}{$\sqrt{M_D M_{\tilde{D}}} \gtrsim 0.52 (0.28)Y_*$ TeV}&\multicolumn{1}{c|}{$\sqrt{M_D M_{\tilde{B}}} \gtrsim 0.75 (0.40)Y_*$ TeV}\\
&\multicolumn{3}{|c|}{}\\[0.01cm]
\cline{1-4}
&\multicolumn{3}{|c|}{}\\[0.02cm]
\multicolumn{1}{|l}{$C^{'CH-UV}_{7}(m_*)$}&\multicolumn{3}{|c|}{$\sim\frac{\left(Y_* v \right)^2 }{M_{D}M_{\tilde{D}}}\xi_{D}\xi_{\tilde{D}}\frac{m_s}{m_b V^{2}_{ts}}$}\\
&\multicolumn{3}{|c|}{}\\[0.01cm]
&\multicolumn{1}{|c}{ESTIMATED} &\multicolumn{1}{c}{TS5}&\multicolumn{1}{c|}{TS10}\\
&\multicolumn{1}{|c}{$\sqrt{M_D M_{\tilde{D}}} \gtrsim (1.1)Y_*$ TeV}&\multicolumn{1}{c}{$\sqrt{M_D M_{\tilde{D}}} \gtrsim (0.40)Y_*$ TeV}&\multicolumn{1}{c|}{$\sqrt{M_D M_{\tilde{B}}} \gtrsim (0.58)Y_*$ TeV}\\
&\multicolumn{3}{|c|}{}\\[0.01cm]
\cline{1-4}
\end{tabular}}}
\caption{\small{Estimated bounds from $b\to s\gamma$ in a generic composite Higgs model and in the specific TS5 and TS10 at small elementary/composite mixing angles $s_1$ and $s_{bR}$. $\xi_{\psi/\chi}$ denotes the degree of compositeness of a SM/Heavy fermion. In the specific TS5 and TS10 models: $\xi_{qL}\equiv s_1$,  $\xi_{D}\equiv c_1$. 
$D=(T,B)$, 
$\tilde{D}$ denotes a $SU(2)_L$ singlet heavy fermion. We highlight (in bold) the MFV bounds from $C^{CH}_7$. For the estimated bounds from $C^{CH}_7$ and for the bounds from $C^{CH-UV}_7$,
 we indicate both the values that can be obtained in the case of a positive (the first number) or a negative 
(the second number in parenthesis) contribution.  
}}
\label{tab:estimates_calc}
\end{table}

\subsection{Non-MFV constraints}\label{non-mfv_bound}
\subsubsection{Generational mixing}\label{generation_mix}
After the EWSB, the mass eigenstate basis is obtained, as in the SM, using unitary transformations: 
$(D_{L}, D_{R})$ and $(U_{L}, U_{R})$ for down and up-type quark respectively. 
We will assume that the left rotation matrix has entries of the same order as
those of the Cabibbo-Kobayashi-Maskawa matrix:
\begin{equation}
 (D_L)_{ij}\sim (V_{CKM})_{ij}\ .
\label{VCKM}
\end{equation}
The assumption of anarchical $Y_*$ fixes the form of the rotation matrix $D_R$ to be:
\begin{equation}
 (D_R)_{ij}\sim \left( \frac{m_i}{m_j}\right)\frac{1}{(D_L)_{ij}}\ \ \text{for}\  i<j\ .
\label{DR}
\end{equation}
\noindent
Considering the estimates (\ref{VCKM}) and (\ref{DR}), 
we can evaluate the generational mixing factors in the composite Higgs model contributions to $C_{7}$ (UV) and $C^{'}_{7}$.\\
For the ultraviolet contribution to $C^{'}_{7}$, we consider the presence of a mass insertion 
that can generate the operator $\bar{b}_{L}\sigma^{\mu\nu}F_{\mu\nu}s_{R}$. 
This mass insertion brings to a factor $m_{b}(D_{R})_{23}\sim\frac{m_{s}}{(D_{L})_{23}}\sim \frac{m_{s}}{V_{ts}}$; 
where we have first used the estimate in (\ref{DR}) and then that in (\ref{VCKM}). 
The ultraviolet contribution to $C_{7}$ involves the operator $\bar{b}_{R}\sigma^{\mu\nu}F_{\mu\nu}s_{L}$ and we obtain, 
from the mass insertion, a 
generational mixing factor $m_b (D_L)_{23}\sim m_b V_{ts}$; where the last similitude follows from the assumption in (\ref{VCKM}).\\
Evaluating, similarly, the generational mixing factor for the vertex $W t_R s_R$ in $C^{'CH-IR}_7$, one finds: 
$(D_R)_{23}\sim\frac{m_s}{m_b (D_L)_{23}}\sim \frac{m_s}{m_b V_{ts}} $, making use, again, of the estimates (\ref{DR}) and (\ref{VCKM}). 
The flavor violation in $C^{CH-IR}_7$ comes entirely from the SM vertex $W t_L s_L$ and it is accounted by a factor $V_{ts}$. 
Therefore, we find that the composite Higgs model contribution to the Wilson coefficient $C^{'}_{7}$ is enhanced by a factor
\begin{equation}
 \frac{m_{s}}{m_{b}V^{2}_{ts}}\sim 8
\label{gen_mix}
\end{equation} 
compared to the contribution to $C_{7}$ both in the ultraviolet and in the infrared case.

\subsubsection{Infrared contribution to $C'_7$}

\begin{figure}
\centering
\scalebox{0.7}{\includegraphics{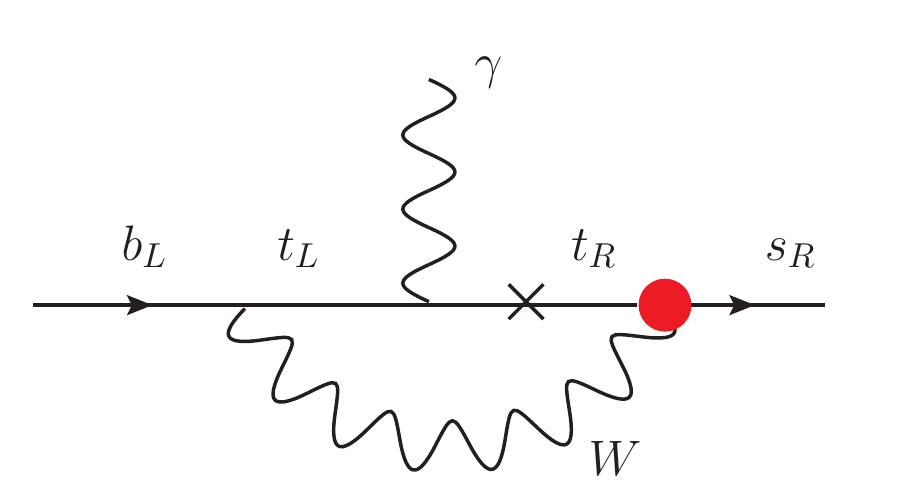}}
\caption{\small{1 loop Infrared contribution to $C'_7$.}}
\label{fig:C7pIR_dia}
\end{figure}

Taking into account the generational mixing factor in (\ref{gen_mix}), 
the composite Higgs model contribution to the Wilson coefficient $C^{'}_{7}$ (in Fig. \ref{fig:C7pIR_dia}) is given by:
\begin{equation}
 C^{'CH-IR}_{7}(\mu_{w})= \frac{\mathcal{C}_{R}v^{2}}{2}\frac{m_{s}}{m_{b}V^{2}_{ts}}\frac{m_{t}}{m_{b}}f_{RH}(x_{t})\ .
\end{equation} 
Considering the estimates for $\mathcal{C}_{R}$ in (\ref{CR_est}) and (\ref{CR_est_pc}), the condition on $C^{'CH-IR}_{7}(\mu_{w})$, eq. (\ref{C'7limIR}), 
gives thus the estimated bounds: 
\begin{equation}
 M_{D}\gtrsim 0.80\ \text{TeV}
\end{equation}
in models with $P_C$ symmetry; and
\begin{equation}
 M_{D}\gtrsim \frac{0.80}{\xi_{qL}} \text{TeV}
\end{equation}
in models without $P_C$ symmetry.\\
\noindent
Considering the specific TS5 and TS10 models, $C^{'CH-IR}_7$ gives the bounds:
\begin{equation}
 M^{TS5}_{D}\gtrsim 1.1\ \text{TeV}
\label{boundC7p_ts5}
\end{equation} 
in the TS5; and
\begin{equation}
 M^{TS10}_{D}\gtrsim\frac{c_{1}}{s_{1}}(0.80)\ \text{TeV}\ 
\label{boundC7p_ts10}
\end{equation} 
in the TS10.

We can discuss how the bound on heavy masses can change in the case of a fully composite top: 
 in the TS5 the bound on doublet heavy fermion (\ref{boundIR-TS5}) does not depend on the top degree of compositeness (this remains
almost true considering the full numerical calculation) and we obtain quite strong MFV bounds both for composite $t_L$ and composite $t_R$. 
In the TS10, because of the $P_C$ protection, we obtain strong bounds in the case of a fully composite $t_R$ (eq. (\ref{boundIR-TS10})). 
Ref. \cite{pomarol_serra} finds that corrections to $S$ and $T$ parameters give only weak constraints on a composite $t_R$ (both in TS5 and in TS10). 
The IR contribution to $b\to s \gamma$, on the contrary, put a quite strong constraint, especially in the TS10, on this limit case.\\
One can finally discuss the validity of our results, which have been obtained `analytically' 
(i.e. by considering an expansion in $x\equiv \frac{Y_{*}v}{\sqrt{2}m_{*}}$ and retaining only the $O(x)$ terms).  
We find that the results from the numerical calculation of the bounds, obtained by diagonalizing numerically the fermionic mass matrices,  
do not differ more than O(1) from those we have shown, which are obtained at order $x$, in the assumption $x\ll 1$.
This can be also found by considering that the
exchange of relatively light custodians, that can give a contribution $\frac{Y_{*}v}{\sqrt{2}m^{CUST}_{*}} > 1$ to the effective $Wt_R b_R$ vertex, 
has to be followed by the exchange of heavier composite fermions, that reduces the overall contribution.
By definition, indeed, the custodians do not directly couple to SM fermions, therefore their contribution to $Wt_R b_R$ is always accompanied by the exchange of heavier composite particles. 

\subsubsection{Ultraviolet contribution} \label{UV_mchm5}

\begin{figure}
\centering
 \includegraphics[width=0.4\textwidth]{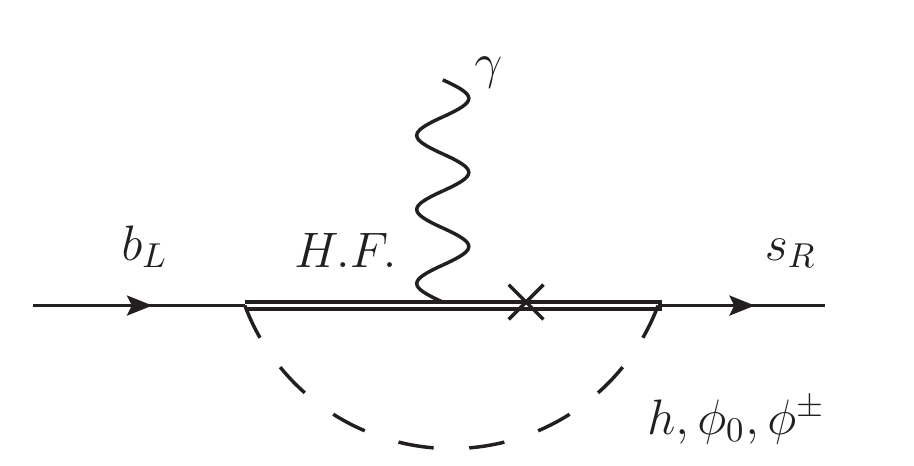}
\caption{\small{1 loop CHM UV contribution to $C^{'}_7$.}}
 \label{UVfig}
\end{figure}

In this case the $P_{C}$ parity does not influence the bounds and we get contributions of the same size in the different models.
The leading contribution comes from diagrams with heavy fermions and would-be Goldstone bosons in the loop\footnote{
The contribution from heavy gluon and heavy fermion exchange is suppressed. 
Indeed this contribution is approximately diagonal in the flavor space.} (Fig. \ref{UVfig}). 

\begin{equation}
C^{CH-UV}_7, C^{'CH-UV}_7 \propto s_{Li}Y_{*ik} Y_{*kl}Y_{*lj}s_{Rj}
 \label{c7UV}
\end{equation} 
The contribution (\ref{c7UV}) is not aligned with the mass matrix $m_{dij}\sim s_{Li} Y_{*ij} s_{Rj}$, therefore,
after the EWSB it remains non diagonal in the flavor space. \\

Before going on the specific TS5 and TS10 models, 
we can obtain estimated bounds from the UV contributions in generic composite Higgs models, by means of NDA.
We obtain:
\begin{equation}
 C^{CH-UV}_7 \sim \frac{\left(Y_* v \right)^2 }{M_{D}M_{\tilde{D}}}\xi_{D}\xi_{\tilde{D}} \ ,
\end{equation}
where $\tilde{D}$ denotes a heavy fermion which is a $SU(2)_L$ singlet, and 
\begin{equation}
 C^{'CH-UV}_7 \sim \frac{m_s}{m_b V^2_{ts}}\frac{\left(Y_* v \right)^2 }{M_{D}M_{\tilde{D}}}\xi_{D}\xi_{\tilde{D}} \ ,
\end{equation}
where we have taken into account the generational mixing factor in (\ref{gen_mix}). 
By comparing these results with those from the IR contributions in (\ref{C7bound_est}, \ref{C7bound_est_pc}), 
we see that the UV contribution gives approximately a bound $Y_{*}/y_t$ ($\frac{Y_{*}}{y_t}\xi_{q_L}$, in the case of models without $P_C$ protection)
times stronger than the one from the IR contribution to $C_7$. 
Such UV bounds, however, are not as robust as the IR one, since they require, as we already pointed out, assumptions on the flavor structure of the BSM sector. 
In particular, we have estimated them in the scenario of flavor anarchy in the strong sector. Notice that in this anarchic scenario much stronger bounds on the resonance masses, of the order of $20$ TeV \cite{Csaki_anarchic}, come from $\epsilon_k$ .\\

\begin{figure}[t]
\title{\textbf{UV contribution in the TS5}}
\includegraphics[width=0.45\textwidth]{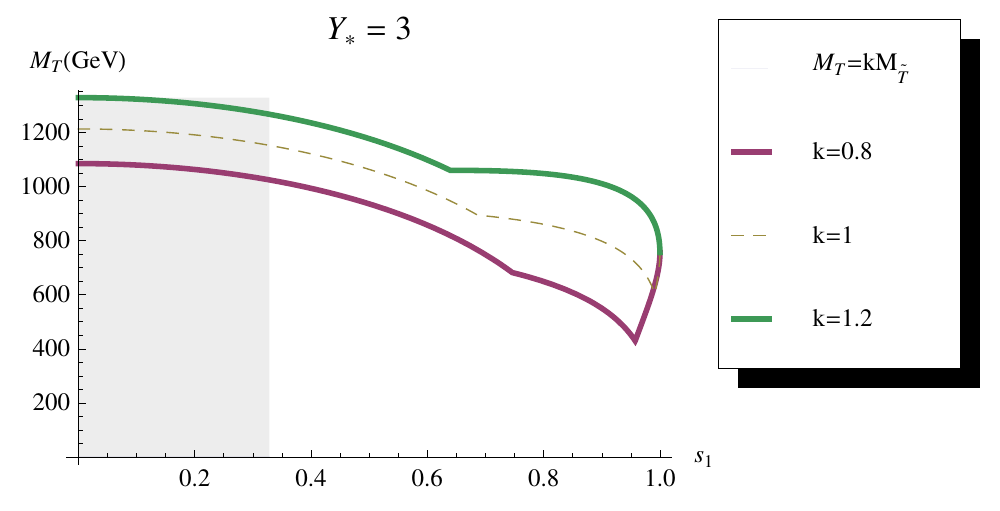}
\includegraphics[width=0.45\textwidth]{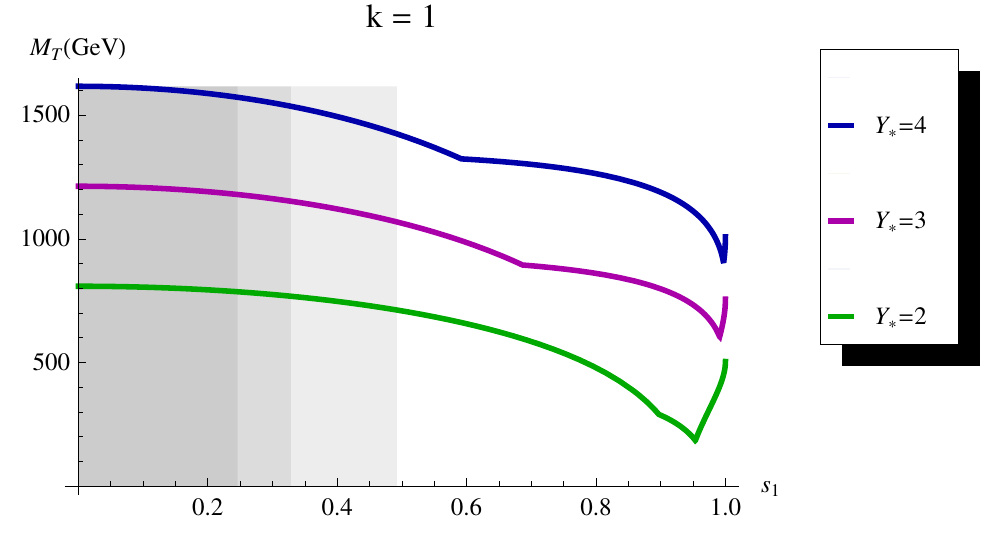}
\caption{\small{ Bounds from $C^{'CH-UV}_7$ in the TS5. Left Plot: bounds for different values of $k=\frac{M_T}{M_{\tilde{T}}}$ and $Y_{*}=3$; Right Plot: bounds for different values of $Y_{*}$ and $k=1$. We set $M_{\tilde{B}}=M_{\tilde{T}}$ and $M_{T'}=M_T$. 
Also shown is the exclusion region for $s_1$, obtained from the condition $s_R=\frac{\sqrt{2} m_t }{Y_* v s_1 }\leq 1$.}}
\label{fig:UV_s1}
\end{figure}

\noindent
In Ref. \cite{Agashe_flav} the Ultraviolet contribution to $b\rightarrow s\gamma$ in a two-site model without a $P_{LR}$ protection to the 
$t_R$ and $b_R$ interactions is evaluated. In the following
 we will describe in details the contribution in the TS5 and we will report the results for TS10.
We can calculate the $C^{CH-UV}_{7}$ and $C^{'CH-UV}_{7}$ ultraviolet contributions by considering
the model independent analysis of Ref. \cite{Agashe_flav} and the generational mixing factor in (\ref{gen_mix}). We get the following effective Hamiltonian for $b\to s \gamma$ with loops of heavy fermions and neutral would-be Goldstone bosons:
\begin{equation}
 \mathcal{H}^{eff}_{neutral\ Higgs}=\frac{i\ e}{8\pi^{2}}\frac{(2\epsilon \cdot p)}{M^{2}_{w}}k_{neutral}\left[V_{ts}\bar{b}(1-\gamma_{5})s+\frac{m_{s}}{m_{b}V_{ts}}\bar{b}(1+\gamma_{5})s \right] 
\label{Heff_dipole}
\end{equation} 
where
\begin{gather}
\nonumber
 k_{neutral}\approx\sum^{4}_{i=1}\left(|\alpha^{(i)}_{1}|^{2}+|\alpha^{(i)}_{2}|^{2}\right)m_{b}\left(\frac{1}{36}\right)\frac{M^{2}_{w}}{m^{2}_{*(i)}} + \\ \nonumber
\sum^{4}_{i=1}\left(\alpha^{(i)*}_{1}\alpha^{(i)}_{2}\right)m_{*(i)}\left(\frac{1}{6}\right)\frac{M^{2}_{w}}{m^{2}_{*(i)}} \\ 
\label{kneutral1}
\end{gather} 
the index $i$ runs over the four down-type heavy fermions of the model, $\mathbf{d}^{(i)}=\tilde{B}, B', B_{-1/3}, B$, 
and the $\alpha^{(i)}_{1}$, $\alpha^{(i)}_{2}$ coefficients are defined by the interactions:
\begin{equation}
 \mathcal{L}\supset \bar{\mathbf{d}}^{(i)}\left[\alpha^{(i)}_{1}(1+\gamma_{5})+\alpha^{(i)}_{2}(1-\gamma_{5})\right]bH + h.c.\ .
\end{equation} 
After the EWSB, we find the following coefficients at $O(x)$: 
\begin{gather}
\nonumber
 \alpha^{(\tilde{B})}_{1}=\frac{Y^2_{*}v}{2}s_{bR}\left[\frac{1}{M_{B'}}+\frac{M_{B'}+c_{bR}M_{\tilde{B}}}{M^2_{B'}-M^2_{\tilde{B}}} \right]\\ \nonumber
 \alpha^{(\tilde{B})}_{2}=-\frac{Y_{*}}{2\sqrt{2}}s_2 c_{bR}\\ \nonumber
 \alpha^{(B')}_{1}=\alpha^{(B_{-1/3})}_{1}=-\frac{Y_{*}}{2\sqrt{2}}s_{bR}\\ \nonumber
 \alpha^{(B')}_{2}=\alpha^{(B_{-1/3})}_{2}=-\frac{Y^2_{*}v}{4}s_{2}\left[ \frac{M^2_{B'}M_{\tilde{B}}-s^2_{bR}M^3_{\tilde{B}}-c_{bR}M^3_{B'}+2c_{bR}M_{B'}M^2_{\tilde{B}}}{M_{B'}M_{\tilde{B}}(M^2_{B'}-M^2_{\tilde{B}})}\right]  \\  
\label{alfa}
\end{gather} 
the heavy fermion $B$ gives a contribution of $O(s^{2}_{2})$ to $k_{neutral}$ and we neglect it.\\
Considering the eq. (\ref{kneutral1}) and the coefficients in (\ref{alfa}), neglecting again $O(x^{2})$ terms, we obtain:
\begin{equation}
  k_{neutral}\approx-m_b M^2_W Y^2_{*}\frac{1}{8}\left(\frac{c_{bR}}{M_{B'}M_{\tilde{B}}}-\frac{7}{18}\frac{s^{2}_{bR}}{M^2_{B'}}\right)   \ .
\label{kneutral}
\end{equation} 
From this expression of $k_{neutral}$ we obtain the following TS5 ultraviolet contributions to the Wilson coefficient of the effective Hamiltonian in (\ref{Heff}):

\begin{align}
\begin{split}
 & C^{CH-UV}_{7}(m_{*})=\frac{1}{16}\frac{\sqrt{2}}{G_{F}}Y^{2}_{*}\left(\frac{c_{bR}}{M_{B'}M_{\tilde{B}}}-\frac{7}{18}\frac{s^{2}_{bR}}{M^2_{B'}}\right)\ ;\\
& C^{'CH-UV}_{7}(m_{*})=\frac{1}{16}\frac{\sqrt{2}}{G_{F}}Y^{2}_{*}\left(\frac{c_{bR}}{M_{B'}M_{\tilde{B}}}-\frac{7}{18}\frac{s^{2}_{bR}}{M^2_{B'}}\right)\frac{m_{s}}{m_{b}V^{2}_{ts}} \ .
\end{split}
\end{align}
Assuming $s_{bR}$ small, the above formulas become:
\begin{equation}
 C^{CH-UV}_{7}(m_{*})=\frac{1}{16}\frac{\sqrt{2}}{G_{F}}\frac{Y^{2}_{*}}{M_{B'}M_{\tilde{B}}}\ ;\ 
C^{'CH-UV}_{7}(m_{*})=\frac{1}{16}\frac{\sqrt{2}}{G_{F}}\frac{Y^{2}_{*}}{M_{B'}M_{\tilde{B}}}\frac{m_{s}}{m_{b}V^{2}_{ts}} \ .
\end{equation}
\noindent
Finally, the condition on $C^{'CH-UV}_{7}$ in the eq. (\ref{C'7lim}) gives the bound:
\begin{equation}
 \sqrt{M_{B'}M_{\tilde{B}}}\gtrsim(0.40)\ Y_{*}\ \text{TeV}\ ;
\label{UVbound}
\end{equation} 
where, for simplicity, we have set $s_{bR}=0$.
The condition (\ref{C7UVlim}) on $C^{CH-UV}_{7}$ gives a stronger bound, 
\begin{equation} 
\sqrt{M_{B'}M_{\tilde{B}}}\gtrsim(0.52)\ Y_{*} \ \text{TeV} \ ,
\end{equation}
 if $C^{CH-UV}_{7}(m_{*})$ is a negative contribution.\\ 
There is also a contribution to $b\rightarrow s\gamma$ from diagrams with heavy fermions and charged Higgs in the loop. 
Following a similar procedure as the one used before (\ref{kchargedApp}) we find, neglecting $O(x^{2})$ terms:
\begin{equation}
  k_{charged}\approx m_b M^2_W Y^2_*\frac{5}{48}\frac{1}{M_{B'}M_{\tilde{B}}}+O(s^2_1)+O(s^2_{bR}) \ .
\end{equation} 
If we can neglect $O(s^{2}_{1})$ and $O(s^{2}_{bR})$ terms, $k_{charged}$ gives a weaker bound than the one from $k_{neutral}$. 
The full expression of $k_{charged}$ can be found in App. \ref{AppUVcontrib}, here we have just reported, for simplicity, the result for small $s_1$ and $s_{bR}$ angles. \\
In Fig. \ref{fig:UV_s1} we show the bound on the doublet mass $M_{T}$ as function of $s_{1}$ from the condition on $C^{'CH-UV}_{7}$, 
for different values of the ratio $k=\frac{M_T}{M_{\tilde{T}}}$ between doublet and singlet masses, fixing $Y_{*}=3$ (Left Plot), and for different value of $Y_{*}$, fixing $k=1$ (Right Plot). We set $M_{\tilde{B}}=M_{\tilde{T}}$ and $M_{T'}=M_T$. These values are obtained by taking into account the strongest values between the neutral Higgs contribution and the charged Higgs one. We set $s_{bR}=s_{1}$.

\subsubsection{Ultraviolet contribution in the TS10}
\begin{figure}[tbp]
\title{\textbf{UV contribution in the TS10}}
\includegraphics[width=0.45\textwidth]{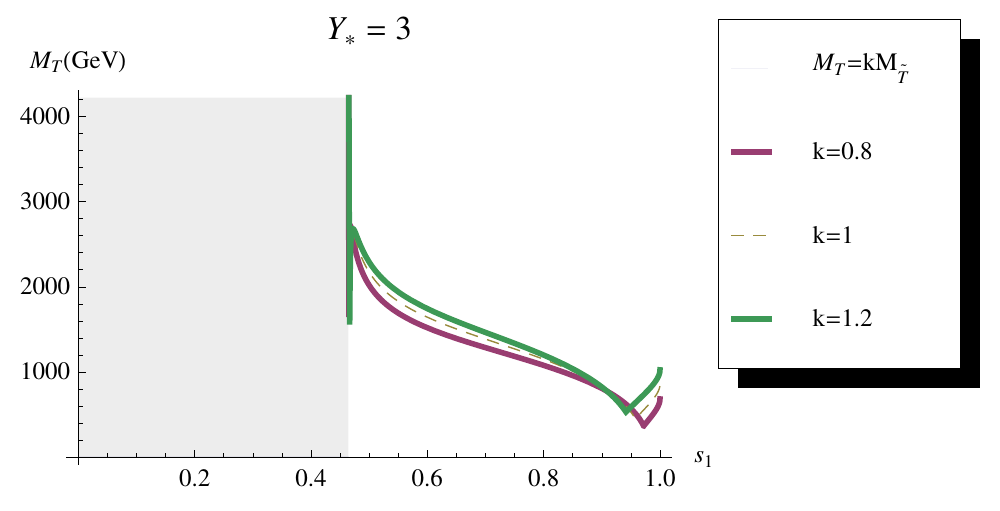}
\includegraphics[width=0.45\textwidth]{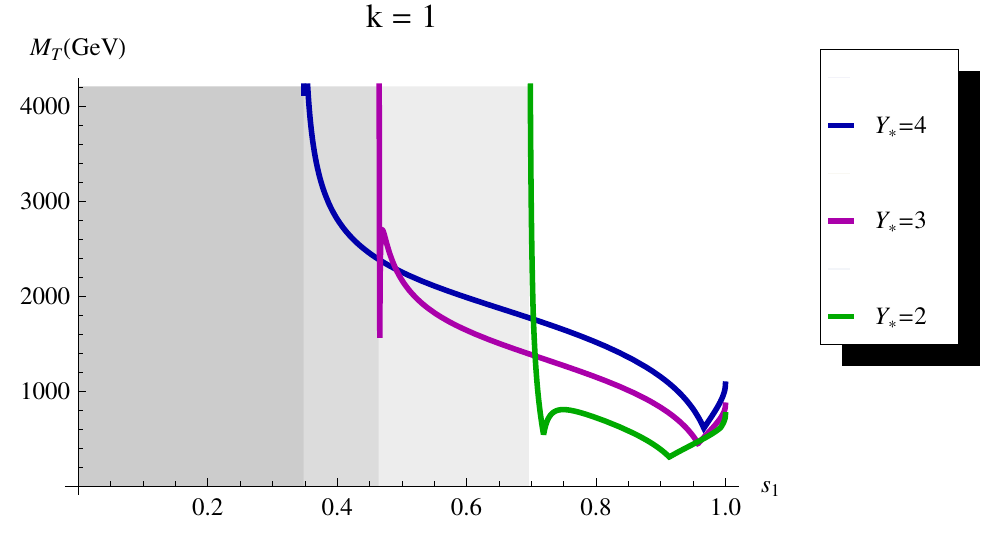}
\caption{\small{ Bounds from $C^{'CH-UV}_7$ in the TS10. Left Plot: bounds for different values of $k=\frac{M_T}{M_{\tilde{T}}}$ ($M_{\tilde{B}}\simeq c_R M_{\tilde{T}}$, $M_{\tilde{B}'}=M_{\tilde{T}'}= c_R M_{\tilde{T}}$), fixing $Y_{*}=3$; Right Plot: bounds for different values of $Y_{*}$, fixing $k=1$. 
We also show the exclusion region for $s_1$, obtained from the condition $s_R=\frac{2 m_t }{Y_* v s_1 }\leq 1$.}}
\label{fig:UV_s1_ts10}
\end{figure}

For the TS10 model, applying the same procedure as for the case of TS5, we get:
\begin{align}
\begin{split}\label{kneutral_ts10}
& k_{neutral} =  m_b M^2_W Y^2_*\\
&\times \frac{7 M_T M^2_{T'} s^2_1 - 18 M_{\tilde{B}} M^2_{\tilde{B}'} \sqrt{1-s^2_1}+ M^2_{\tilde{B}} \left( 7 M_B s^2_1 -18 M_{\tilde{B}'} \sqrt{1-s^2_1}\right)}{288 M^2_{\tilde{B}} M_B M^2_{\tilde{B}'}} +O(s_{bR})\\
 & = - m_b M^2_W Y^2_{*}\frac{1}{16}\left( \frac{1}{M_{B}M_{\tilde{B}}}+\frac{1}{M_{B}M_{\tilde{B}'}}\right) +O(s^2_1)+O(s_{bR})
\end{split}
\end{align}

\begin{equation}
 k_{charged}= m_b M^2_W Y^2_* \left(\frac{5}{48}\frac{1}{M_{B}M_{\tilde{B}}}+\frac{5}{48}\frac{1}{M_{B}M_{\tilde{B}'}}+ \frac{5}{96}\frac{s^2_R}{M^2_{B}}\right) +O(s^2_1)+O(s^2_{bR})
\label{kcharged_ts10}
\end{equation} 
If the left-handed bottom quark has a small degree of compositeness, we can neglect $O(s^{2}_{1})$ 
(while $s_{bR}$ is naturally very small in the TS10, in order to account for the ratio $m_b/m_t\ll 1$). 
The charged contribution, in this case, gives a stronger bound than the one from $k_{neutral}$: 
\begin{equation}
 \sqrt{M_{B}M_{\tilde{B}}}\gtrsim(0.58)\ Y_{*}\ \text{TeV}\ ,
\label{UVbound-ts10}
\end{equation} 
from the condition (\ref{C'7lim}) on $C^{'CH-UV}_{7}$. A stronger bound, 
\begin{equation}
\sqrt{M_{B}M_{\tilde{B}}}\gtrsim(0.75)\ Y_{*} \ \text{TeV} \ ,
\end{equation} 
comes from the condition (\ref{C7UVlim}) on $C^{CH-UV}_{7}$, if this last contribution has a negative sign.\\
\noindent
In Fig. \ref{fig:UV_s1_ts10} we show the bound on the doublet mass $M_{T}$ as function of $s_{1}$ from the condition on $C^{'CH-UV}_{7}$, 
for different values of the ratio $k=\frac{M_T}{M_{\tilde{T}}}$ between doublet and $\tilde{T}$ singlet mass, fixing $Y_{*}=3$ (Left Plot), and for different $Y_{*}$ values, setting $k=\frac{M_T}{M_{\tilde{T}}}=1$ (Right Plot).
The custodian singlet masses have the following relations with $M_{\tilde{T}}$:
$M_{\tilde{B}}\simeq c_R M_{\tilde{T}}$, $M_{\tilde{B}'}=M_{\tilde{T}'}= c_R M_{\tilde{T}}$.
All these bounds are obtained by taking into account the strongest values between the neutral Higgs contribution and the charged Higgs one.\\ 
We can see that in the TS10 model, the UV bounds are particularly strong in the case of fully composite $t_R$. This is an effect caused by the exchange of the custodians
$\tilde{T}'$, $\tilde{B}'$ and of the $\tilde{B}$, that are light in the limit of a composite $t_R$. 
In particular, when $t_R$ is fully composite ($s_R=1$), 
$M_{\tilde{B}}(\simeq c_R M_{\tilde{T}})$ and $M_{\tilde{B}'}=M_{\tilde{T}'}(= c_R M_{\tilde{T}})$ vanish. 
This causes the divergence of the bounds for $s_R \to 1$. 
Such divergences can be seen in the curves in Figure \ref{fig:UV_s1_ts10}, 
when they approach the (grey) exclusion regions for $s_1$ (indeed, the minimum value of $s_1$ 
allowed by the condition $s_R=\frac{2 m_t }{Y_* v s_1 }\leq 1$ is obviously obtained in the case $s_R=1$).\\

%

\noindent
Tab. \ref{tab:estimates_calc} summarizes our results. It shows the bounds on heavy fermion masses that can be obtained from the process $b\to s \gamma$. 
We report the estimated bounds in generic Composite Higgs Models (with or without $P_C$ protection), which we have found by means of NDA, 
and the bounds in the specific two-site models TS5 and TS10. 
$\xi_{\psi/\chi}$ denotes the degree of compositeness of a SM/Heavy fermion. In the specific TS5 and TS10 models: $\xi_{qL}\equiv s_1$,  $\xi_{D}\equiv c_1$. 
$D=(T,B)$, 
$\tilde{D}$ denotes a $SU(2)_L$ singlet heavy fermion. For the estimated bounds from $C^{CH}_7$ and for the bounds from $C^{CH-UV}_7$,
 we indicate both the values that can be obtained in the case of a positive (the first number) or a negative 
(the second number in parenthesis) contribution.

\subsection{Constraint from $\epsilon^{'}/\epsilon_{K}$}
The bound on the mass of the heavy fermions that comes from the direct CP violating observable
of the $K^0\to 2\pi$ system, $Re(\epsilon '/\epsilon)$, can be even stronger, in the assumption of anarchic $Y_*$, 
than those obtained from $b\to s \gamma$, as already found in \cite{Isidori}. 
As we pointed out, however, it is a bound that strongly depends on the assumptions made on the flavor structure of the new physics sector.\\ 
As for the UV contribution to $b\rightarrow s\gamma$, the custodial symmetry does not influence the bound and we obtain contributions of the same size 
in the different models. In what follows we describe the bound in the TS5 and in the TS10.\\ 
\noindent
New Physics contribution can be parametrized at low energy by chromo-magnetic operators:
\begin{equation}
 \mathcal{O}_{G}=\bar{s}\sigma^{\mu\nu}T^{a}G^{a}_{\mu\nu}\left(1-\gamma_{5} \right)d \ \ , \ \  \mathcal{O'}_{G}=\bar{s}\sigma^{\mu\nu}T^{a}G^{a}_{\mu\nu}\left(1+\gamma_{5} \right)d\ .
\end{equation} 
As for the UV contribution to $b\rightarrow s\gamma$, the leading contribution to $\epsilon^{'}/\epsilon_{K}$ comes from diagrams with heavy fermions and Higgs in the loop, 
that generate the $\mathcal{O}_{G}$ and $\mathcal{O'}_{G}$ operators (1 loop diagrams are the same as for the UV contribution to $b\to s \gamma$, Fig. \ref{UVfig}, 
with the replacements $\gamma \to g$, $b \to s$ and $s \to d$). \\
The related coefficients $\mathcal{C}_{G}$ and $\mathcal{C'}_{G}$, in analogy with $\mathcal{C}_{7}$ and $\mathcal{C'}_{7}$ of the UV contribution to $b\rightarrow s\gamma$,
differ by a generational mixing factor that, in the assumption of anarchic $Y_{*}$, we estimate to be $\sim \frac{m_{d}}{m_{s}V^{2}_{us}}$. 
We consider only the generation mixing $(1-3)\times (2-3)$, via 3rd generation.\\
In analogy with (\ref{Heff_dipole}), we define:
\begin{equation}
 \mathcal{A}^{eff-chromo}_{neutral\ Higgs}=\frac{i\ g_{s}}{8\pi^{2}}\frac{(2\epsilon \cdot p)}{M^{2}_{w}}k^{G}_{neutral}\left[V_{us}\bar{s}(1-\gamma_{5})d+\frac{m_{d}}{m_{s}V_{us}}\bar{s}(1+\gamma_{5})d \right]\ , 
\label{Heff_chromo}
\end{equation} 
where
\begin{gather}
\nonumber
 k^{G}_{neutral}\approx\sum^{4}_{i=1}\left(|\alpha^{(i)}_{1}|^{2}+|\alpha^{(i)}_{2}|^{2}\right)m_{s}\left(-\frac{1}{12}\right)\frac{M^{2}_{w}}{m^{2}_{*(i)}} + \\ \nonumber
\sum^{4}_{i=1}\left(\alpha^{(i)*}_{1}\alpha^{(i)}_{2}\right)m_{*(i)}\left(-\frac{1}{2}\right)\frac{M^{2}_{w}}{m^{2}_{*(i)}} \\ 
\label{kGneutral1}
\end{gather} 
the index $i$ runs over the four down-type heavy fermions of the model, $\mathbf{d}^{(i)}$, 
and the $\alpha^{(i)}_{1}$, $\alpha^{(i)}_{2}$ coefficients are defined by the interactions:
\begin{equation}
 \mathcal{L}\supset \bar{\mathbf{d}}^{(i)}\left[\alpha^{(i)}_{1}(1+\gamma_{5})+\alpha^{(i)}_{2}(1-\gamma_{5})\right]bH + h.c.\ .
\end{equation} 
\noindent
After the EWSB, neglecting $O(x^{2})$ terms, we find in the TS5:
\begin{equation}
  k^{G}_{neutral}= \frac{3}{8}m_{s}M^{2}_{w}\frac{Y^{2}_{*}}{M_{B'}M_{\tilde{B}}} + O(s^2_{sR})\ ,
\label{kGneutral2}
\end{equation} 
\noindent
where $s_{sR}$ defines the degree of compositeness of the right-handed strange quark and has naturally a small value. 
In the limit in which $s_{sR}=0$, we obtain the same result also in the TS10. \\
\noindent
We can thus calculate the $\mathcal{C}_{G}$ and $\mathcal{C'}_{G}$ contributions:
\begin{equation}
 \mathcal{C}_{G}=-\frac{1}{16 \pi^2}\frac{k^{G}_{neutral}}{M^{2}_{w}m_{s}}V_{us} \ , \ \mathcal{C'}_{G}=\frac{m_{d}}{m_{s}V^{2}_{us}}\mathcal{C}_{G}\ .
\end{equation} 
\noindent
Defining
\begin{equation}
 \delta_{\epsilon^{'}}=\frac{Re(\epsilon^{'}/\epsilon)_{CH}-Re(\epsilon^{'}/\epsilon)_{SM}}{Re(\epsilon^{'}/\epsilon)_{exp}}
\end{equation} 
\noindent
we obtain
\begin{equation}
 |\delta_{\epsilon^{'}}|\approx \left( 58\ TeV\right)^{2}B_{G}|\mathcal{C}_{G}-\mathcal{C'}_{G}|<1 \ ,
\label{epsPbound}
\end{equation} 
\noindent
where $Re(\epsilon^{'}/\epsilon)_{SM}$ has been estimated as in Ref. \cite{Isidori}; $B_G$ denotes the hadronic bag-parameter, 
$\left\langle 2\pi_{I=0}| y_s \mathcal{O}_{G} | K^0 \right\rangle $. 
We take $B_{G}=1$\footnote{ That corresponds to the estimate of the hadronic matrix element $\left\langle 2\pi_{I=0}| y_s \mathcal{O}_{G} | K^0 \right\rangle $
 in the chiral quark
model and to the first order in the chiral expansion.} and we take into account separately the contribution from $\mathcal{C}_{G}$ and $\mathcal{C'}_{G}$.\\
In the limit $s_{sR}=0$ we obtain from (\ref{epsPbound}):

\begin{equation}
 \sqrt{M_{B'}M_{\tilde{B}}} \gtrsim  (1.3) Y_*\  TeV\ ,
\label{epsPbound_2}
\end{equation} 
\noindent
which is in agreement with the result in \cite{Isidori}.
The contribution from the charged Higgs interactions gives weaker bounds than those from the neutral Higgs contribution.

\section{Conclusions}
\label{sec:conclusions}

Composite Higgs Models are among the compelling scenarios for physics beyond  
the Standard Model that can give an explanation of the origin of the EWSB
and that are going to be tested at the LHC.\\
In this project we have have built simple ``two-site" models, the TS5 and the TS10, 
which can represent the low energy regime of Minimal Composite Higgs Models with a custodial symmetry and a $P_{LR}$ parity.\\
Working in these effective descriptions, we have reconsidered the bounds on the CHM spectrum implied by flavor observables.
We have found in particular that the IR contribution to $b\to s \gamma$ induced by the flavor conserving effective vertex $W t_R b_R$ 
implies a robust Minimal Flavor Violating bound on the mass ($m_*$) of the new heavy fermions (to be more specific, on the heavy doublets, partners of $q_L=(t_L,b_L)$). 
The relevance of shifts to $W t_R b_R$ has been already pointed out in the literature (see, for example, \cite{Misiak2008, Drobnak2012}), 
even though its importance in setting a bound on heavy fermion masses was unestimated in previous studies. 
We have also shown how this bound can be stronger in the case of the absence of a symmetry ($P_C$) protection to the effective $Wt_R b_R$ vertex. 
In particular, we have found an estimated bound
\[
m_{*}\gtrsim 1.0\ \text{TeV} \ ,
\]
in models with $P_C$ protection to the $Wt_R b_R$ vertex (where both $t_R$ and $b_R$ are $P_C$ eigenstates) and a bound
\[
m_{*}\gtrsim 1.0/\xi_{qL}\ \text{TeV}\ , 
\]
where $\xi_{qL}$ denotes the degree of compositeness of $(t_L,b_L)$, in models without $P_C$ protection. $\xi_{qL}$ is naturally a small number, the bound could be thus very strong in these types of models. In the specific ``two-site" models, the bounds we have found are
\[m^{TS5}_* \gtrsim 1.4\ \text{TeV}
\]
in the TS5, and
\[m^{TS10}_* \gtrsim \frac{0.54}{\xi_{qL}}\ \text{TeV}\ , 
\]
in the TS10.\\
Table \ref{tab:estimates_calc} summarizes the results obtained for the bounds from $b\to s \gamma$.\\
In addition to these bounds, we have calculated the constraints from the UV composite Higgs model contribution to $b \to s \gamma$. Figs. \ref{fig:UV_s1} and \ref{fig:UV_s1_ts10} show the bounds in the TS5 and the TS10 as functions of the $t_L$ degree of compositeness.
Our results have shown that these bounds can be stronger than those from the IR contribution but they are model dependent; 
in particular they strongly depend on the assumptions made on the flavor structure of the composite sector. 
We have obtained an estimated limit
 \[
m_{*}\gtrsim (0.52)Y_{*}\ \text{TeV}
 \]
in a specific NP flavor scenario ($Y_{*}$ anarchic in the flavor space). \\ 
Even stronger bounds,
\[
m_{*}\gtrsim (1.3)Y_{*}\ \text{TeV} \ ,
\] 
can be obtained from $\epsilon^{'}/\epsilon_K$ 
but, again, they are model dependent and in principle could be loosened by acting on the NP flavor structure (as done, for example, in Ref. \cite{Redi_Weiler}).
The lower IR bounds on $m_*$ we have found from $b\to s \gamma$, on the contrary, are robust MFV bounds that cannot be evaded by assuming particular conditions on the structure of the strong sector. 

\section*{Acknowledgments}
I would like to thank Roberto Contino for having followed this work from the beginning and for comments on the manuscript.

\appendix
\section*{Appendix}
\section{Two Site Models}
\subsection{TS5}
\label{TS5A}

Fermions rotate from the elementary/composite basis to the `physical' light(SM)/heavy basis as
(we neglect $O\left(\Delta^{2}_{L2}\right)$ terms):

\begin{align}
\begin{split}\label{rotation_LL}
& \tan\varphi_{L1}=\frac{\Delta_{L1}}{M_{Q*}}\equiv \frac{s_{1}}{c_{1}}, \ \  s_{1}\equiv\sin\varphi_{L1} \  c_{1}\equiv\cos\varphi_{L1}\\
& s_{2}=\frac{\Delta_{L2}}{M_{Q'*}}\cos\varphi_{L1}\\ 
& s_{3}=\frac{\Delta_{L2}M_{Q'*}}{\Delta^{2}_{L1}+M^{2}_{Q*}-M^{2}_{Q'*}}\sin\varphi_{L1} \\
& \left\{\begin{array}{l}
	t_{L}=c_{1}t^{el}_{L}-s_{1}T^{com}_{L}-s_{2}T'^{com}_{L}\\
	T_{L}=s_{1}t^{el}_{L}+c_{1}T^{com}_{L}+s_{3}T'^{com}_{L} \\
        T'_{L}=\left(s_{2}c_{1}-s_{1}s_{3}\right)t^{el}_{L}-\left(s_{1}s_{2}+c_{1}s_{3}\right)T^{com}_{L}+ T'^{com}_{L} 
\end{array}  \right. \\
& \left\{\begin{array}{l}
	b_{L}=c_{1}b^{el}_{L}-s_{1}B^{com}_{L}-s_{2}B'^{com}_{L}\\
	B_{L}=s_{1}b^{el}_{L}+c_{1}B^{com}_{L}+s_{3}B'^{com}_{L} \\
        B'_{L}=\left(s_{2}c_{1}-s_{1}s_{3}\right)b^{el}_{L}-\left(c_{1}s_{3}+s_{1}s_{2}\right)B^{com}_{L}+B'^{com}_{L}
\end{array} \right.
\end{split}
\end{align}

\begin{align}
\begin{split}\label{rotation_LR}
& s_{4}=\Delta_{L2}\frac{\Delta_{L1}}{\Delta^{2}_{L1}+M^{2}_{Q*}-M^{2}_{Q'*}} \\
& \left\{\begin{array}{l}
	T_{R}=T^{com}_{R}+s_{4}T'^{com}_{R}\\
	T'_{R}=T'^{com}_{R}-s_{4}T^{com}_{R}
\end{array}  \right.  \ \ 
 \left\{\begin{array}{l}
	B_{R}=B^{com}_{R}+s_{4}B'^{com}_{R}\\
	B'_{R}=B'^{com}_{R}-s_{4}B^{com}_{R}  
\end{array} \right.
\end{split}
\end{align}

\begin{align}
\begin{split}\label{rotation_RR}
& \tan\varphi_{R}=\frac{\Delta_{R1}}{M_{\tilde{T}*}}\ \ s_{R}\equiv\sin\varphi_{R} \ \ c_{R}\equiv\cos\varphi_{R} \\
& \tan\varphi_{bR}=\frac{\Delta_{R2}}{M_{\tilde{B}*}}\ \ s_{bR}\equiv\sin\varphi_{bR} \ \ c_{bR}\equiv\cos\varphi_{bR} \\
& \left\{\begin{array}{l}
	t_{R}=c_{R}t^{el}_{R}-s_{R}\tilde{T}^{com}_{R}\\
	\tilde{T}_{R}=s_{R}t^{el}_{R}+c_{R}\tilde{T}^{com}_{R} 
\end{array}  \right. \ \ 
 \left\{\begin{array}{l}
	b_{R}=c_{bR}b^{el}_{R}-s_{bR}\tilde{B}^{com}_{R} \\ 
	\tilde{B}_{R}=s_{bR}b^{el}_{R}+c_{bR}\tilde{B}^{com}_{R}
\end{array} \right.
\end{split}
\end{align}\\

\noindent
Physical heavy fermion masses are related to the bare ones according to:
\begin{align}
\left\{\begin{array}{l}
M_{\tilde{T}}=\sqrt{M^{2}_{\tilde{T}*}+\Delta^{2}_{R1}}=\frac{M_{\tilde{T}*}}{c_{R}}\\
M_{\tilde{B}}=\sqrt{M^{2}_{\tilde{B}*}+\Delta^{2}_{R2}}=\frac{M_{\tilde{B}*}}{c_{bR}}\\
M_{T}=M_{B}=\sqrt{M^{2}_{Q*}+\Delta^{2}_{L1}}=\frac{M_{Q*}}{c_{1}} \\
M_{T5/3}=M_{T2/3}=M_{Q*} \\
M_{T'}=M_{B'}=\sqrt{M^{2}_{Q'*}+\Delta^{2}_{L2}}\simeq M_{Q'*}=M_{B-1/3}=M_{B-4/3} \end{array} \right.
\end{align}\\

\noindent
In the elementary/composite basis the Yukawa Lagrangian reads: 
\begin{align}
\begin{split}	
\mathcal{L}^{YUK} =& \ Y_{*U}Tr\left\{\bar{\mathcal{Q}}\mathcal{H}\right\}\tilde{T}+Y_{*D}Tr\left\{\bar{\mathcal{Q'}}\mathcal{H}\right\}\tilde{B}+h.c.\\ 
=& \ Y_{*U}\left\{\bar{T}\phi^{\dag}_{0}\tilde{T}+\bar{T}_{2/3}\phi_{0}\tilde{T}+\bar{T}_{5/3}\phi^{+}\tilde{T}-\bar{B}\phi^{-}\tilde{T}\right\}\\ 
& +Y_{*D}\left\{\bar{B}_{-1/3}\phi^{\dag}_{0}\tilde{B}+\bar{B'}\phi_{0}\tilde{B}+\bar{T'}\phi^{+}\tilde{B}-\bar{B}_{-4/3}\phi^{-}\tilde{B}\right\}+h.c.
\end{split}
\end{align}
\noindent
After field rotation to the mass eigenstate basis, before EWSB, $\mathcal{L}^{YUK}$ reads as in eq. (\ref{eq.Lagrange2}).\\

\noindent
After the EWSB top and bottom masses arise as:
\begin{equation}
	m_{t}=\frac{v}{\sqrt{2}}Y_{*U}s_{1}s_{R}
\label{tmass}
\end{equation}
\begin{equation}
	m_{b}=\frac{v}{\sqrt{2}}Y_{*D}s_{2}s_{bR} \ .
\label{bmass}
\end{equation}
\noindent
We have also electroweak mixings among fermions. The fermionic mass matrices for up and down states read, 
in the basis $\left(\bar{t}_L\ \bar{\tilde{T}}_L\ \bar{T}_{2/3L}\ \bar{T}_L\ \bar{T'}_L \right)$ 
$\left(t_R\ \tilde{T}_R\ T_{2/3R}\ T_R\ T'_R \right)$ for the up sector 
and in the basis $\left(\bar{b}_L\ \bar{\tilde{B}}_L\ \bar{B'}_L\ \bar{B}_{-1/3L} \ \bar{B}_L\right)$ 
$\left(b_R\ \tilde{B}_R\ B'_R\ B_{-1/3R} \ B_R\right)$ for the down-type fermions:

\begin{align}
\begin{split}\label{Mup}
&\mathcal{M}_{up}= \\
& \left(\begin{array}{ccccc}
	m_{t} & -Y_{*U}\frac{v}{\sqrt{2}}s_{1}c_{R} & 0 & 0& 0\\ 
	0 & M_{\tilde{T}} & Y_{*U}\frac{v}{\sqrt{2}} & Y_{*U}\frac{v}{\sqrt{2}}& -s_{4}Y_{*U}\frac{v}{\sqrt{2}}\\ 
 -Y_{*U}\frac{v}{\sqrt{2}}s_{R}& Y_{*U}\frac{v}{\sqrt{2}}c_{R} & M_{T2/3} & 0& 0\\ 
	-Y_{*U}\frac{v}{\sqrt{2}}c_{1}s_{R} &Y_{*U}\frac{v}{\sqrt{2}}c_{1}c_{R} & 0 & M_T & 0\\ 
	Y_{*U}\frac{v}{\sqrt{2}}\left(s_{1}s_{2}+c_{1}s_{3}\right)s_{R} & -Y_{*U}\frac{v}{\sqrt{2}}\left(s_{1}s_{2}+c_{1}s_{3}\right)c_{R}& 0& 0& M_{T'}
	\end{array}\right)
\end{split}
\end{align} 

\begin{align}
\begin{split}\label{Mdown}
&\mathcal{M}_{down}=\\ 
& \left(\begin{array}{ccccc}
	m_{b} & -Y_{*D}\frac{v}{\sqrt{2}}s_{2}c_{bR} & 0 & 0 & 0\\
	0 & M_{\tilde{B}} & Y_{*D}\frac{v}{\sqrt{2}} & Y_{*D}\frac{v}{\sqrt{2}} & Y_{*D}\frac{v}{\sqrt{2}}s_{4} \\
	-Y_{*D}\frac{v}{\sqrt{2}}s_{bR} & Y_{*D}\frac{v}{\sqrt{2}}c_{bR} & M_{B'} & 0 & 0\\
	-Y_{*D}\frac{v}{\sqrt{2}}s_{bR} & 	Y_{*D}\frac{v}{\sqrt{2}}c_{bR} & 0 & M_{B-1/3} & 0\\
		-Y_{*D}\frac{v}{\sqrt{2}}s_{3}s_{bR} & 	Y_{*D}\frac{v}{\sqrt{2}}s_{3}c_{bR} & 0 & 0 & M_B\\
	\end{array}\right)
\end{split}
\end{align} 

\begin{align}
\begin{split}\label{eq.Lagrange2}	
\mathcal{L}^{YUK}= & Y_{*U}c_{1}c_{R}\left(\bar{T}_{L}\phi^{\dag}_{0}\tilde{T}_{R}-\bar{B}_{L}\phi^{-}\tilde{T}_{R}\right)+Y_{*U}c_{R}\left(\bar{T}_{2/3L}\phi_{0}\tilde{T}_{R}+\bar{T}_{5/3L}\phi^{+}\tilde{T}_{R}\right)\\ 
& -Y_{*U}\left(s_{1}s_{2}+c_{1}s_{3}\right)c_{R}\left(\bar{T'}_{L}\phi^{\dag}_{0}\tilde{T}_{R}-\bar{B'}_{L}\phi^{-}\tilde{T}_{R}\right)-
 Y_{*U}s_{1}c_{R}\left(\bar{t}_{L}\phi^{\dag}_{0}\tilde{T}_{R}-\bar{b}_{L}\phi^{-}\tilde{T}_{R}\right)\\ 
& -Y_{*U}s_{R}\left(\bar{T}_{2/3L}\phi_{0}t_{R}+\bar{T}_{5/3L}\phi^{+}t_{R}\right)+Y_{*U}\left(s_{1}s_{2}+c_{1}s_{3}\right)s_{R}\left(\bar{T'}_{L}\phi^{\dag}_{0}t_{R}-\bar{B'}_{L}\phi^{-}t_{R}\right)\\ 
& -Y_{*U}c_{1}s_{R}\left(\bar{T}_{L}\phi^{\dag}_{0}t_{R}-\bar{B}_{L}\phi^{-}t_{R}\right)+Y_{*U}s_{1}s_{R}\left(\bar{t}_{L}\phi^{\dag}_{0}t_{R}-\bar{b}_{L}\phi^{-}t_{R}\right)\\ 
&+Y_{*U}\left(\bar{T}_{R}\phi^{\dag}_{0}\tilde{T}_{L}-\bar{B}_{R}\phi^{-}\tilde{T}_{L}\right)+Y_{*U}\left(\bar{T}_{2/3R}\phi_{0}\tilde{T}_{L}+\bar{T}_{5/3R}\phi^{+}\tilde{T}_{L}\right)\\ 
&-Y_{*U}s_{4}\left(\bar{T'}_{R}\phi^{\dag}_{0}\tilde{T}_{L}-\bar{B'}_{R}\phi^{-}\tilde{T}_{L}\right)\\ 
&+Y_{*D}c_{bR}\left(\bar{B}_{-1/3L}\phi^{\dag}_{0}\tilde{B}_{R}-\bar{B}_{-4/3L}\phi^{-}\tilde{B}_{R}\right)+Y_{*D}c_{bR}\left(\bar{B'}_{L}\phi_{0}\tilde{B}_{R}+\bar{T'}_{L}\phi^{+}\tilde{B}_{R}\right)\\ 
&-Y_{*D}s_{bR}\left(\bar{B}_{-1/3L}\phi^{\dag}_{0}b_{R}-\bar{B}_{-4/3L}\phi^{-}b_{R}\right)-Y_{*D}s_{bR}\left(\bar{B'}_{L}\phi_{0}b_{R}+\bar{T'}_{L}\phi^{+}b_{R}\right)\\ 
&-Y_{*D}s_{2}c_{bR}\left(\bar{b}_{L}\phi_{0}\tilde{B}_{R}+\bar{t}_{L}\phi^{+}\tilde{B}_{R}\right)+Y_{*D}s_{2}s_{bR}\left(\bar{b}_{L}\phi_{0}b_{R}+\bar{t}_{L}\phi^{+}b_{R}\right)\\ 
&-Y_{*D}s_{3}s_{bR}\left(\bar{B}_{L}\phi_{0}b_{R}+\bar{T}_{L}\phi^{+}b_{R}\right)+Y_{*D}s_{3}c_{bR}\left(\bar{B}_{L}\phi_{0}\tilde{B}_{R}+\bar{T}_{L}\phi^{+}\tilde{B}_{R}\right)\\
&+Y_{*D}\left(\bar{B'}_{R}\phi_{0}\tilde{B}_{L}+\bar{T'}_{R}\phi^{+}\tilde{B}_{L}\right)+Y_{*U}\left(\bar{B}_{-1/3R}\phi^{\dag}_{0}\tilde{B}_{L}-\bar{B}_{-4/3R}\phi^{-}\tilde{B}_{L}\right)\\ 
&+Y_{*D}s_{4}\left(\bar{B}_{R}\phi_{0}\tilde{B}_{L}+\bar{T}_{R}\phi^{+}\tilde{B}_{L}\right)+h.c. 
\end{split}
\end{align}

\subsection{TS10}
\label{TS10A}

Fermions rotate from the elementary/composite basis to the 'physical' light(SM)/heavy basis as:
\begin{align}
\begin{split}\label{rotation_LL_TS}
& \tan\varphi_{L1}=\frac{\Delta_{L1}}{M_{Q*}}\equiv \frac{s_{1}}{c_{1}}\\  
& \left\{\begin{array}{l}
	t_{L}=c_{1}t^{el}_{L}-s_{1}T^{com}_{L}\\
	T_{L}=s_{1}t^{el}_{L}+c_{1}T^{com}_{L}
\end{array} \right.  \ \ 
 \left\{\begin{array}{l}
	b_{L}=c_{1}b^{el}_{L}-s_{1}B^{com}_{L}\\
	B_{L}=s_{1}b^{el}_{L}+c_{1}B^{com}_{L}  
\end{array} \right.
\end{split}
\end{align}

\begin{align}
\begin{split}\label{rotation_RR_TS}
& \tan\varphi_{R}=\frac{\Delta_{R1}}{M_{\tilde{Q}*}}\ \ s_{R}\equiv\sin\varphi_{R} \ \ c_{R}\equiv\cos\varphi_{R} \\
& \tan\varphi_{bR}=\frac{\Delta_{R2}}{M_{\tilde{Q}*}}\ \ s_{bR}\equiv\sin\varphi_{bR} \ \ c_{bR}\equiv\cos\varphi_{bR} \\
& \left\{\begin{array}{l}
	t_{R}=c_{R}t^{el}_{R}-s_{R}\tilde{T}^{com}_{R}\\
	\tilde{T}_{R}=s_{R}t^{el}_{R}+c_{R}\tilde{T}^{com}_{R} 
\end{array}  \right.  \ \ 
 \left\{\begin{array}{l}
	b_{R}=c_{bR}b^{el}_{R}-s_{bR}\tilde{B}^{com}_{R} \\ 
	\tilde{B}_{R}=s_{bR}b^{el}_{R}+c_{bR}\tilde{B}^{com}_{R}
\end{array} \right. 
\end{split}
\end{align}\\

\noindent
Physical heavy fermion masses are related to the bare ones as:
\begin{align}
\left\{\begin{array}{l}
M_{\tilde{T}}=\sqrt{M^{2}_{\tilde{Q}*}+\Delta^{2}_{R1}}=\frac{M_{\tilde{Q}*}}{c_{R}}\\
M_{\tilde{B}}=\sqrt{M^{2}_{\tilde{Q}*}+\Delta^{2}_{R2}}=\frac{M_{\tilde{Q}*}}{c_{bR}}\\
M_{\tilde{T}5/3}=M_{\tilde{T}'5/3}=M_{\tilde{T}'}=M_{\tilde{B}'}=M_{\tilde{Q}*}\\
M_{T}=M_B=\sqrt{M^{2}_{Q*}+\Delta^{2}_{L1}}=\frac{M_{Q*}}{c_{1}} \\
M_{T2/3}=M_{T5/3}=M_{Q*} 
\end{array} \right.
\end{align}\\

\noindent
In the elementary/composite basis the Yukawa Lagrangian reads:
\begin{equation}	
\mathcal{L}^{YUK}=+Y_{*}Tr\left\{\mathcal{H}\bar{\mathcal{Q}}\mathcal{\tilde{Q}'}\right\}+Y_{*}Tr\left\{\bar{\mathcal{Q}}\mathcal{H}\mathcal{\tilde{Q}}\right\} 
\end{equation}
\noindent

After field rotation to the mass eigenstate basis, before EWSB, $\mathcal{L}^{YUK}$ reads as in eq. (\ref{eq.Lagrange2_ts10}).\\

\noindent
After EWSB top and bottom masses arise as:

\begin{equation}
	m_{t}=\frac{v}{2}Y_{*}s_{1}s_{R}
\label{tmass_TS}
\end{equation}
\begin{equation}
	m_{b}=\frac{v}{\sqrt{2}}Y_{*}s_{1}s_{bR}
\label{bmass_TS}
\end{equation}\\

\noindent
The fermionic mass matrices for up and down states read, 
in the basis $\left(\bar{t}_L\ \bar{\tilde{T}}_L\ \bar{T}_{2/3L}\ \bar{T}_L\ \bar{\tilde{T}}'_L \right)$ 
$\left(t_R\ \tilde{T}_R\ T_{2/3R}\ T_R\ \tilde{T}'_R \right)$ for the up sector 
and in the basis $\left(\bar{b}_L\ \bar{\tilde{B}}_L\ \bar{\tilde{B}}'_L\ \bar{B}_L\right)$ 
$\left(b_R\ \tilde{B}_R\ \tilde{B}'_R\ B_R\right)$ for the down-type fermions:

\begin{align}
\mathcal{M}^{TS10}_{up}=\ Y_{*}\frac{v}{2}\left(\begin{array}{ccccc}
	\frac{m_{t}}{Y_{*}\frac{v}{2}} & -s_{1}c_{R} & 0 & 0& -s_{1}\\ 
	0 & \frac{M_{\tilde{T}}}{Y_{*}\frac{v}{2}} & -1& 1& 0\\ 
 s_{R} & -c_{R} & \frac{M_{T2/3}}{Y_{*}\frac{v}{2}} & 0& -1\\ 
	-c_{1}s_{R} & c_{1}c_{R} & 0 & \frac{M_{T}}{Y_{*}\frac{v}{2}} & c_{1}\\ 
	0 & 0 & -1& 1& \frac{M_{\tilde{T}'}}{Y_{*}\frac{v}{2}}
	\end{array}\right)
\label{Mup_mchm10}
\end{align}

\begin{align}
\mathcal{M}^{TS10}_{down}=\ Y_{*}\frac{v}{\sqrt{2}}\left(\begin{array}{cccc}
	\frac{m_{b}}{Y_{*}\frac{v}{\sqrt{2}}} & -s_{1}c_{bR} & -s_{1} & 0 \\
	0 & \frac{M_{\tilde{B}}}{Y_{*}\frac{v}{\sqrt{2}}} & 0 & 1\\
	0 & 0 &\frac{ M_{\tilde{B}'}}{Y_{*}\frac{v}{\sqrt{2}}} & 1 \\
	-c_{1}s_{bR} & 	c_{1}c_{bR} & c_{1} & \frac{M_B}{Y_{*}\frac{v}{\sqrt{2}}}\\
	\end{array}\right)
\label{Mdown_mchm10}
\end{align}

\begin{align}
\begin{split}\label{eq.Lagrange2_ts10}	
\mathcal{L}^{YUK}=\  & Y_{*}c_{1}c_{R}\frac{1}{\sqrt{2}}\left(\bar{T}_{L}\phi^{\dag}_{0}\tilde{T}_{R}-\bar{B}_{L}\phi^{-}\tilde{T}_{R}\right)-Y_{*}c_{R}\frac{1}{\sqrt{2}}\left(\bar{T}_{2/3L}\phi_{0}\tilde{T}_{R}+\bar{T}_{5/3L}\phi^{+}\tilde{T}_{R}\right)\\ 
& - Y_{*}s_{1}c_{R}\frac{1}{\sqrt{2}}\left(\bar{t}_{L}\phi^{\dag}_{0}\tilde{T}_{R}-\bar{b}_{L}\phi^{-}\tilde{T}_{R}\right)+Y_{*}s_{1}s_{R}\frac{1}{\sqrt{2}}\left(\bar{t}_{L}\phi^{\dag}_{0}t_{R}-\bar{b}_{L}\phi^{-}t_{R}\right)\\ 
& +Y_{*}s_{R}\frac{1}{\sqrt{2}}\left(\bar{T}_{2/3L}\phi_{0}t_{R}+\bar{T}_{5/3L}\phi^{+}t_{R}\right)
 -Y_{*}c_{1}s_{R}\frac{1}{\sqrt{2}}\left(\bar{T}_{L}\phi^{\dag}_{0}t_{R}-\bar{B}_{L}\phi^{-}t_{R}\right)\\ 
& +Y_{*}\frac{1}{\sqrt{2}}\left(\bar{T}_{R}\phi^{\dag}_{0}\tilde{T}_{L}-\bar{B}_{R}\phi^{-}\tilde{T}_{L}\right)-Y_{*}\frac{1}{\sqrt{2}}\left(\bar{T}_{2/3R}\phi_{0}\tilde{T}_{L}+\bar{T}_{5/3R}\phi^{+}\tilde{T}_{L}\right)\\ 
& +Y_{*}\left(\bar{T}_{5/3L}\phi^{\dag}_{0}\tilde{T}_{5/3R}-\bar{T}_{2/3L}\phi^{-}\tilde{T}_{5/3R}\right)+Y_{*}\left(\bar{T}_{5/3R}\phi^{\dag}_{0}\tilde{T}_{5/3L}- \bar{T}_{2/3R}\phi^{-}\tilde{T}_{5/3L}\right)\\
&-Y_{*}s_{1}c_{bR}\left(\bar{b}_{L}\phi_{0}\tilde{B}_{R}+\bar{t}_{L}\phi^{+}\tilde{B}_{R}\right)+Y_{*}s_{1}s_{bR}\left(\bar{b}_{L}\phi_{0}b_{R}+\bar{t}_{L}\phi^{+}b_{R}\right)\\ 
&-Y_{*}c_{1}s_{bR}\left(\bar{B}_{L}\phi_{0}b_{R}+\bar{T}_{L}\phi^{+}b_{R}\right)+Y_{*}c_{1}c_{bR}\left(\bar{B}_{L}\phi_{0}\tilde{B}_{R}+\bar{T}_{L}\phi^{+}\tilde{B}_{R}\right)\\
&+Y_{*}\left(\bar{B}_{R}\phi_{0}\tilde{B}_{L}+\bar{T}_{R}\phi^{+}\tilde{B}_{L}\right)+ Y_{*}\left(\bar{B}_{R}\phi^{\dag}_{0}\tilde{B}'_{L}+Y_{*}\bar{T}_{2/3R}\phi^{+}\tilde{B}'_{L}\right)\\
& Y_{*}\frac{1}{\sqrt{2}}\left(\bar{T}_{R}\phi^{\dag}_{0}\tilde{T}'_{L}+\bar{B}_{R}\phi^{-}\tilde{T}'_{L}\right)-Y_{*}\frac{1}{\sqrt{2}}\left(\bar{T}_{2/3R}\phi_{0}\tilde{T}'_{L}-\bar{T}_{5/3R}\phi^{+}\tilde{T}'_{L}\right)\\
& + Y_{*}c_{1}\frac{1}{\sqrt{2}}\left(\bar{T}_{L}\phi^{\dag}_{0}\tilde{T}'_{R}+\bar{B}_{L}\phi^{-}\tilde{T}'_{R}\right)-Y_{*}\frac{1}{\sqrt{2}}\left(\bar{T}_{2/3L}\phi^{\dag}_0\tilde{T}'_{R}-\bar{T}_{5/3L}\phi^{+}\tilde{T}'_{R}\right)\\
&- Y_{*}s_{1}\frac{1}{\sqrt{2}}\left(\bar{t}_{L}\phi^{\dag}_{0}\tilde{T}'_{R}+\bar{b}_{L}\phi^{-}\tilde{T}'_{R}\right)+Y_{*}\left(\bar{T}_{5/3R}\phi_{0}\tilde{T}'_{5/3L}-\bar{T}_{R}\phi^{-}\tilde{T}'_{5/3L}\right)\\
&+Y_{*}c_{1}\left(\bar{B}_{L}\phi^{\dag}_{0}\tilde{B}'_{R}-\bar{T}_{L}\phi^{-}\tilde{T}'_{5/3R}\right)-Y_{*}s_{1}\left(\bar{b}_{L}\phi^{\dag}_{0}\tilde{B}'_{R}-\bar{t}_{L}\phi^{-}\tilde{T}'_{5/3R}\right)\\
&+Y_{*}\bar{T}_{2/3L}\phi^{+}\tilde{B}'_{R}+Y_* \bar{T}_{5/3L}\phi_{0}\tilde{T}'_{5/3R} + h.c.
\end{split}
\end{align}

\section{BOUND derivation}
\label{App_bound}
The SM prediction and the experimental measurement \cite{BsGamma_exp} of the $b\to s \gamma$ branching ratio are respectively:
\begin{equation}
 BR_{th}=(315 \pm 23)10^{-6}
 \label{BRth}
\end{equation}
 \begin{equation}
  BR_{ex}=(355 \pm 24 \pm 9)10^{-6}
 \label{BRex}
\end{equation}
The $b\to s \gamma$ decay rate is:
\begin{equation}
\Gamma_{tot} \propto |\mathcal{C}_7(\mu_b)|^2+|\mathcal{C}^{'}_7(\mu_b)|^2\approx 
|\mathcal{C}^{SM}_7(\mu_b)+\mathcal{C}^{NP}_7(\mu_b)|^2+|\mathcal{C}^{'NP}_7(\mu_b)|^2
\label{gamma_tot}
\end{equation}
\noindent
If we consider only the $\mathcal{C}_7$ contribution, we obtain:
\begin{equation}
\frac{\Gamma_{tot}}{\Gamma_{SM}}=1+2\frac{Re(\mathcal{C}^{SM}_7(\mu_b)^{*}\mathcal{C}^{NP}_7(\mu_b))}{|\mathcal{C}^{SM}_7(\mu_b)|^2}+O(\Delta\mathcal{C}^2_7)
\label{gamma_ratio}
\end{equation}
For $\mu_b =5$ GeV, $\mu_W = M_W $, $\alpha_S =0.118$, the SM contribution to $C_7$ at the scale $\mu_b$ reads \cite{Buras}:

\begin{equation}
C^{SM}_7(\mu_b)=0.695 C^{SM}_7(\mu_W)+0.086 C^{SM}_8(\mu_W)-0.158 C^{SM}_2(\mu_W)=-0.300 \ .
\label{C7SM}
\end{equation}
The scaling factor of the NP contribution to $C_7$ from the scale $\mu_W$ to the scale $\mu_b$ is:

\begin{equation}
 \mathcal{C}^{NP}_7(\mu_b) = \left(\frac{\alpha_S (\mu_W)}{\alpha_S (\mu_b)} \right)^{\frac{16}{23}}\mathcal{C}^{NP}_7(\mu_w)= 0.695\ \mathcal{C}^{NP}_7(\mu_w) \ .
\label{C7NPscale}
\end{equation}
\noindent
By considering all the previous equations, we obtain at $95 \%$ C.L.:
\[
 -0.0775 < \mathcal{C}^{NP}_7(\mu_w) < 0.0226 \ .
\]
\noindent
The scaling factor of the NP contribution to $C_7$ from the scale $m_* = 1$ TeV  to the scale $\mu_W$ is:
\begin{equation}
 \mathcal{C}^{NP}_7(\mu_W)=\left(\frac{\alpha_S (m_*)}{\alpha_S (m_t)} \right)^{\frac{16}{21}} \left(\frac{\alpha_S (m_t)}{\alpha_S (\mu_W)} \right)^{\frac{16}{23}}\simeq 0.79\ \mathcal{C}^{NP}_7(m_*) \
 \label{scale_mStar}
\end{equation}
\noindent
and we obtain at $95 \%$ C.L.:
\[
 -0.0978 < \mathcal{C}^{NP}_7(m_*) < 0.0284
\]
\noindent 
If we consider only the $\mathcal{C}^{'}_7$ contribution, we obtain:

\begin{equation}
 \frac{\Gamma_{tot}}{\Gamma_{SM}} \simeq 1+\frac{|\mathcal{C}^{'NP}_7(\mu_b)|^2}{|\mathcal{C}^{SM}_7(\mu_b)|^2} \ .
\label{C7pratio}
\end{equation}
\noindent
We have
\begin{equation}
 C^{'}_7(\mu_b)\simeq C^{'NP}_7(\mu_b)=\left(\frac{\alpha_S (m_*)}{\alpha_S (m_t)} \right)^{\frac{16}{21}} \left(\frac{\alpha_S (m_t)}{\alpha_S (\mu_b)} \right)^{\frac{16}{23}} \mathcal{C}^{'NP}_7(m_*)\simeq 0.55\  \mathcal{C}^{'NP}_7(m_*) \ .
 \label{C7p_scale}
\end{equation}
\noindent
By considering (\ref{BRth}), (\ref{BRex}), (\ref{scale_mStar}), (\ref{C7pratio}), (\ref{C7p_scale}), we obtain at $95 \% $ C.L.:

\[
 |\mathcal{C}^{'NP}_7(\mu_w)|<0.294
\]
\[
 |\mathcal{C}^{'NP}_7(m_*)|<0.372
\]




\section{Charged Higgs ultraviolet contribution to $b\rightarrow s\gamma$ in the TS5}
\label{kchargedApp}

\begin{equation}
 \mathcal{H}^{eff}_{charged\ Higgs}=\frac{i\ e}{8\pi^{2}}\frac{(2\epsilon \cdot p)}{M^{2}_{w}}k_{charged}\left[V_{ts}\bar{b}(1-\gamma_{5})s+\frac{m_{s}}{m_{b}V_{ts}}\bar{b}(1+\gamma_{5})s \right] 
\end{equation} 
where
\begin{gather}
\nonumber
 k_{charged}\approx\sum^{4}_{i=1}\left(|\alpha^{(i)}_{1}|^{2}+|\alpha^{(i)}_{2}|^{2}\right)m_{b}\left(-\frac{2}{9}\right)\frac{M^{2}_{w}}{m^{2}_{*(i)}} + \\ \nonumber
\sum^{4}_{i=1}\left(\alpha^{(i)*}_{1}\alpha^{(i)}_{2}\right)m_{*(i)}\left(-\frac{5}{6}\right)\frac{M^{2}_{w}}{m^{2}_{*(i)}} \\ 
\label{kcharged}
\end{gather} 
the index $i$ runs over the four up-type heavy fermions of the model, $\mathbf{u}^{(i)}$, $m_{*(i)}$ denotes the physical mass of the the $\mathbf{u}^{(i)}$ heavy fermion
and the $\alpha^{(i)}_{1}$, $\alpha^{(i)}_{2}$ coefficients derive from the interactions:
\begin{equation}
 \mathcal{L}\supset \bar{\mathbf{u}}^{(i)}\left[\alpha^{(i)}_{1}(1+\gamma_{5})+\alpha^{(i)}_{2}(1-\gamma_{5})\right]bH^{+} + h.c.\ .
\end{equation} \\

After the EWSB, we diagonalize the up-type quarks mass matrix of (\ref{Mup}) and the down-type one (\ref{Mdown}) perturbatively in $x\equiv\left(\frac{Y_{*}v}{\sqrt{2}m_{*}}\right)$, 
neglecting $O(x^{2})$. We find the following coefficients:\\

\begin{footnotesize}
\begin{gather}
\nonumber
 \alpha^{(\tilde{T})}_{1}= v Y^2_{*} s_1 s_2 s_{bR} \frac{M^2_T M^3_{\tilde{T}} + M^2_{T'} M^3_{\tilde{T}}-M^5_{\tilde{T}}+c_R M^3_T M^2_{T'} c_1}{4 M_T M_{\tilde{T}}(M^2_T-M^2_{\tilde{T}})(-M^2_{T'} + M^2_{\tilde{T}})c_1} \\ \nonumber
 \alpha^{(\tilde{T})}_{2}=\frac{Y_{*}s_1 c_R}{2 \sqrt{2}}\\ \nonumber
 \alpha^{(T)}_{1}= \frac{Y_{*}s_1 s_2 s_{bR}M^2_{T'}}{2 \sqrt{2}c_1 (M^2_{T'}-M^2_T)}\\ \nonumber
 \alpha^{(T)}_{2}= \frac{v Y^2_{*} s_1}{4}\left(\frac{c_R M_{\tilde{T}}+c_1 c^2_R M_T}{M^2_T-M^2_{\tilde{T}}}+\frac{c_1 s^2_R}{M_T}\right) \\  \nonumber
\alpha^{(T')}_{1}=-\frac{s_{bR}Y_{*}}{2 \sqrt{2}}\\ \nonumber
 \alpha^{(T')}_{2}= \frac{Y^2_{*}v s_2}{4}\frac{s^2_1 c_R M_{\tilde{B}}M_{\tilde{T}}M_T M^2_{T'}+s^2_1 c_1 c^2_R M_{\tilde{B}}M^2_T M^2_{T'}+c_1(M^2_{T'}-M^2_{\tilde{T}})\left(M^3_{T'}c_{bR}+M^2_T(s^2_1 s^2_R M_{\tilde{B}} -c_{bR}M_{T'}) \right) }{c_1 M_{\tilde{B}}M_{T'}(M^2_{T'}-M^2_T)(M^2_{T'}-M^2_{\tilde{T}})} \\ 
\label{alfaCharged}
\end{gather} 
\end{footnotesize}
the heavy fermion $T_{2/3}$ 
gives a contribution of $O(x^{2})$ to $k_{charged}$ and we can neglect it.\\
Considering the eq.(\ref{kcharged}) and the coefficients in (\ref{alfaCharged}), neglecting again $O(x^{2})$ terms, we obtain:
\[ k_{charged}=\]
\begin{footnotesize}
\[
 -m_b M^2_W Y^2_*\frac{-15 M^2_T M_{T'} M^2_{\tilde{T}}\sqrt{1-s^2_{bR}} + M_{\tilde{B}} (15 M^2_{T'} M^2_{\tilde{T}} s^2_1 s^2_R + M^2_T (11 M^2_{T'}s^2_1 (-1 + s^2_R) + M^2_{\tilde{T}} (4 s^2_{bR} + 15 s^2_1 s^2_R)))}{144 M_{\tilde{B}} M^2_T M^2_{T'} M^2_{\tilde{T}}} \]
\end{footnotesize}

\normalsize
and:
 \begin{equation}
  k_{charged}\approx m_b M^2_W Y^2_*\frac{5}{48}\frac{1}{M_{B'}M_{\tilde{B}}}+O(s^2_1)+O(s^2_{bR}) \ ,
\label{kcharged_app2}
\end{equation} 
if we can neglect $O(s^{2}_{1})$.

\section{Ultraviolet contribution}\label{AppUVcontrib}
Summing up, we find in the TS5:
\[
k_{neutral} = -m_b M^2_W Y^2_{*}\frac{1}{8}\left(\frac{c_{bR}}{M_{B'}M_{\tilde{B}}}-\frac{7}{18}\frac{s^{2}_{bR}}{M^2_{B'}}\right)= - m_b M^2_W Y^2_{*}\frac{1}{8}\frac{1}{M_{B'}M_{\tilde{B}}}+O(s^2_{bR})
\]

\[k_{charged}=\]

\begin{footnotesize}\[-m_b M^2_W Y^2_*\frac{-15 M^2_T M_{T'} M^2_{\tilde{T}}\sqrt{1-s^2_{bR}} + M_{\tilde{B}} (15 M^2_{T'} M^2_{\tilde{T}} s^2_1 s^2_R + M^2_T (11 M^2_{T'}s^2_1 (-1 + s^2_R) + M^2_{\tilde{T}} (4 s^2_{bR} + 15 s^2_1 s^2_R)))}{144 M_{\tilde{B}} M^2_T M^2_{T'} M^2_{\tilde{T}}}\]\end{footnotesize}
\[
=m_b M^2_W Y^2_*\frac{5}{48}\frac{1}{M_{B'}M_{\tilde{B}}}+O(s^2_1)+O(s^2_{bR})
\]
\noindent
and in the TS10:
\begin{align}
\begin{split}
  k_{neutral} & = m_b M^2_W Y^2_* \frac{7 M_T M^2_{T'} s^2_1 - 18 M_{\tilde{B}} M^2_{\tilde{B}'} \sqrt{1-s^2_1} + M^2_{\tilde{B}} (7 M_B s^2_1 -18 M_{\tilde{B}'} \sqrt{1-s^2_1})}{288 M^2_{\tilde{B}} M_B M^2_{\tilde{B}'}}+O(s_{bR})\\[0.4cm] 
& = - m_b M^2_W Y^2_{*}\frac{1}{16}\left( \frac{1}{M_{B}M_{\tilde{B}}}+\frac{1}{M_{B}M_{\tilde{B}'}}\right) +O(s^2_1)+O(s_{bR})\\[0.7cm]
 k_{charged} & = m_b M^2_W Y^2_* \left(\frac{5}{48}\frac{1}{M_{B}M_{\tilde{B}}}+\frac{5}{48}\frac{1}{M_{B}M_{\tilde{B}'}}+ \frac{5}{96}\frac{s^2_R}{M^2_{B}}\right) +O(s^2_1)+O(s^2_{bR})
 \end{split}
\end{align}


\end{document}